\documentstyle[preprint,12pt,aps,prabib]{revtex}

\begin{document}

\draft

\title{ {\bf The $pd \leftrightarrow \pi^+ t$ reaction
 around the $\Delta$ resonance}}

\author{L. Canton$^1$ and W. Schadow$^2$} 
\address{$^1$Istituto Nazionale di Fisica Nucleare, Sezione di Padova,
Via F. Marzolo 8, I-35131 Padova, Italy}
\address{$^2$Physikalisches
Institut der Universit\"at Bonn, Endenicher Allee 11-13, D-53115
 Bonn, Germany}

\maketitle

\begin{abstract} 

The $pd \leftrightarrow \pi^+ t$
process has been calculated in the energy region
around the $\Delta$--resonance with elementary
production/absorption mechanisms 
involving one and two nucleons.
The isobar degrees of freedom have been 
explicitly included in the two--nucleon mechanism 
via $\pi$-- and $\rho$--exchange diagrams.
No free parameters have been employed in the analysis since
all the parameters 
have been fixed in previous 
studies on the simpler $pp\leftrightarrow \pi^+ d$
process.
The treatment of the few--nucleon dynamics entailed 
a Faddeev--based calculation of the reaction, with 
continuum calculations for the initial $p$--$d$ state and
accurate solutions of the three--nucleon bound--state equation.
The integral cross--section was found to be 
quite sensitive to the NN interaction employed
while the angular dependence showed less sensitivity.
Approximately a 4\% effect was found
for the one--body mechanism, for the
three--nucleon dynamics in the $p$--$d$ channel,
and for the inclusion of a large, possibly converged,
number of three--body partial states,
indicating that these different aspects
are of comparable importance in the calculation
of the spin--averaged observables.
\end{abstract}


\narrowtext

\section{Introduction}

Pion production/absorption on nuclear systems
represents a complex, challenging problem, and this fact has been 
known since many years.
In the attempt to explain the multitude of experimental results 
collected over more than 40 years by now,
many different theoretical approaches have been proposed so far,
with the aim to improve our understanding of these reactions
and, more generally, of the hadronic phenomena.

The simplest approaches which have been employed
assume a one--nucleon mechanism originated by the $\pi$NN vertex
(generally -- but not always -- a Galilei invariant
non relativistic reduction of the usual $\gamma_5$  
pion--nucleon interaction)
and recast the transition amplitude in a DWIA formalism where the 
many--body aspects can be treated with different 
levels of approximations \cite{eise-koltu}.
First--order corrections from the $\pi$-nucleus
multiple scattering series leads to the two--nucleon mechanisms
where a pion emitted from one nucleon scatters from another
before leaving the whole nucleus.

The non--perturbative character of the hadronic interactions, 
together with the energy--momentum mismatch which forbids
absorption on single, free nucleons and suppresses 
one--nucleon absorption on bound nucleons,
makes these rescattering effects an important aspect
which cannot be ignored for the reproduction
of low--energy data. At the $\Delta$--resonance, however,
most of the assumptions used in the early calculations
were not applicable, and other phenomenological
approaches have been developed.

A relevant success was encountered by the ``deuteron" model,
originally employed by Ruderman\cite{ruderman},
where the $pd\rightarrow\pi^+t$ cross section is directly linked
to the $pp\rightarrow\pi^+ d$ experimental cross section,
by means of suitable nuclear structure functions.
This method has the considerable advantage to disentangle the question
of the choice of the correct interaction mechanisms from the
knowledge of the bound state wavefunctions.
The theoretical uncertainties about the interaction mechanisms
at the resonance are simply bypassed through the employment of the
experimental $pp\rightarrow \pi^+ d$ cross section.
The approach has been later refined by several authors, which soon 
addressed their attention to the main issues such as
the role of the distortion effects, sensitivity of the results
with respect to the available model wave--functions,
and generalization of the formalism from three--baryon systems
to the many--nucleon $A(p,\pi^+)A+1$ case.
A partial but representative sample of works developed along these lines
is given in Refs. \cite{ingram,fearing1,LochWeb}.

More recently, the analysis of the $A(p,\pi^+)A+1$ reaction
in terms of the $pp\rightarrow\pi^+d$ process has been extended to spin 
observables by Falk \cite{falk}, by using the $pp\rightarrow\pi^+d$ 
amplitude analysis of Bugg {\it et al.} \cite{bugg},
rather than the averaged cross--section data of earlier works.
In most cases, an overall fair, qualitative agreement with
the trend of the large bulk of experimental data
has been found, indicating 
that the deuteron model even in its spin--dependent version may be considered a
starting point for phenomenological studies
which include spin--polarization data.
This kind of approaches, however, suffers from a number of problems
which limits possible future refinements and demands
for more appropriate theoretical formulations.
First, the model considers NN production mechanisms limited to the 
(dominant) $1 \rightarrow 0$ isospin transition,
while a complete model should take into account also the effects in the 
remaining isospin channels. Second, the amplitudes used for input are
on--shell quantities, while in the theory the off--shell effects should be 
taken into account.
Third, there is a certain arbitrariness in the kinematical constraints
which define the energy parameter for the evaluation of the
input 2N production amplitude; because of that, the results are not 
uniquely determined. And finally, in practically all 
applications the distortion 
contributions are in danger of double countings; indeed, in the 
evaluation of the distortion effects one should subtract the 
distortion contributions on the active 
pair, which are already included in the $pp\rightarrow\pi^+d$ 
data.

To overcome these limitations, 
and to challenge the validity of the ``deuteron" approaches,
there has been a number of papers where
more microscopic models have been 
suggested\cite{gremaq,gresai,sainio}.
Here, the dynamical input was not mediated or hidden by the
cross--section data, but consists of non--relativistic interactions
among pions, nucleons, and isobars. In particular,
the coupled $\Delta$N dynamics is obtained through the solution
in R--space of the two--baryon Schr\"odinger equation.
First calculations\cite{gremaq} employed simple s--wave deuteron and triton
wave--functions of uncorrelated gaussian forms, and were 
extremely limited in the number of intermediate states
included in the calculation. Subsequent analyses 
\cite{gresai,sainio}
used better parameterizations (including $d$ states)
for the nuclear wave--functions, and increased
slightly (from 1 to 4 or 5)
the number of two--body angular--decomposed intermediate states, finding
that these aspects improve the calculated angular distributions
without, however, finding an agreement with theory and experiments,
especially for the results at large angles, where considerable
discrepancies persist. One of the conclusions from these studies was that
these microscopic calculations needed important improvements
in order to give a good reproduction in the normalization
and in the angular distributions, and that
these improvements would lead to enormous complications
in the theoretical evaluation of the observables.

In an attempt to go beyond the description of meson production
in terms of single--nucleon and two--nucleon mechanisms
\cite{lagelecol},
it was shown that meson double rescattering could be a good candidate to 
account for the discrepancies at backward angles, particularly in the 
energy region above the $\Delta$--resonance.
However, the bulk results, while moving towards the right
direction, were still far from being optimal and this
again poses the question of the need to overcome the various 
technical approximations which had to be assumed to keep the calculation 
tractable. The study has the merit to address the
attention to the problem of three--body mechanisms in meson 
production/absorption, and this is still an open question
which we hope will be theoretically disentangled in the near future
with the help of the results collected in recent years
by pionabsorption experiments with large angular
detector systems
\cite{backetal1,backetal2,backetal3} 
and by making comparative analyses in the $\Delta$ region
with the phenomenology of $^3$He photodisintegration \cite{audit}.

In this paper, we have calculated the excitation function 
and differential cross--section for the $p d\rightarrow\pi t$
reaction in the $\Delta$ region by using single--nucleon
and two--nucleon mechanisms. In particular, the two--nucleon mechanism
explicitly refers to the intermediate isobar excitation with
$\pi$+$\rho$ exchange $\Delta$N--NN transitions
with tensor components. The one--body mechanism
is mediated by the $\pi$NN vertex, while the two--body process is
triggered by the $\pi\Delta$N one.
The two mechanisms have been decomposed in complete three--nucleon
partial waves, while the asymptotic pion--nucleus
plane wave has been kept three--dimensional.  
This choice leads to matrix elements with 
a large number of couplings between different three--body states
and the technical complications involved 
has been kept under control with a great deal of numerical
analysis. Also the intermediate three--baryon ($\Delta$NN)
state has been represented in partial waves with inclusion
of angular momenta up to $l=2$ for the $\Delta$N
subsystem. However, in preliminary test calculations
\cite{can95}, the quality of the convergence
has been checked up to $l=5$.
The antisymmetrization prescription
for the three nucleons has been fully taken into account
via the formalism of the permutation operators
\cite{casveca} and brings into the theory 
further mechanisms which differ from the
leading ones by exchange diagrams.  
The matrix elements of the 
symmetrization permutator $P$
involve a large number of couplings between different
three--nucleon partial waves, and rather than
invoking a drastical reduction of the number of states,
we have accepted to deal with the numerical difficulties
implied by the approach.

Following the formalism developed in Ref. \cite{casveca}
we have embedded these absorption operators
in a Faddeev--based treatment of the three--nucleon dynamics.
Accurate bound--state wavefunctions have
been obtained with high--rank Faddeev--AGS calculations,
and a suitable generalization of the quasiparticle method
to absorption processes has been formulated
in order to take into account the three--nucleon dynamics
in the $pd$ channel. This continuum calculation
has been performed in  the rank--one (separable) approximation
and therefore is not as accurate as the bound--state
wavefunction, but this is the first calculation which, to our knowledge,
includes a Faddeev--based treatment of the initial--state
three--body dynamics for the $pd\rightarrow\pi+ t$ process
(and the other processes related by isospin symmetry).

Because of the above mentioned reasons, we think that the treatment 
illustrated in this work 
achieves several improvements with respect to the previous, 
pioneering, microscopic analyses. 
We have devoted section II to further discussions
on these and other important aspects of our calculations,
such as the treatment of the $\Delta$ mass and width
in the intermediate three--body ($\Delta$NN) Green's function,
and the inclusion of off--shell effects
(for this second aspect, see also the discussion at the
beginning of Sect. III).
The remaining part of Sect. III compares the results of our
analysis with experiments and Sect. IV contains a brief
summary and the conclusions.

\section{Theory}

We express the transition amplitude
for the $\pi^+ t \leftrightarrow pd$ reaction
by the following matrix element

\begin{equation}
A_d^{\text{tot}} = {_{\cal S}\langle}\Psi_d^{(-)}|
{\cal A}|\Psi_{\text{BS}}\rangle_{\cal S} \,
|{\bf P}^{\pi}_0\rangle \,,
\end{equation}

\noindent
where $|\Psi_{\text{BS}}\rangle_{\cal S}$ represents 
the three--nucleon bound--state (BS) 
and $_{\cal S}\langle\Psi_d^{(-)}|$ the three--body 
continuum state with ingoing boundary condition
and with the deuteron $d$ in the asymptotic channel.
Both states are assumed to be properly antisymmetrized
in the nucleonic coordinates and the state $|{\bf P}^{\pi}_0\rangle$
is the pion--nucleus three--dimensional plane wave.
In a previous exploratory calculation \cite{can95}
the bound--state wavefunction calculated 
in Ref. \cite{sauer} was used. That wavefunction 
originally was determined in Ref.\cite{sauer}
by solving the Faddeev equation
with the Paris potential and with explicit allowance
of the $\Delta$ degrees of freedom and consisted of 48
three--baryon partial waves, of which 22 refer to purely nucleonic 
states. Because of the smallness of the remaining isobar states,
only the nucleonic states were retained in the calculation
of Ref.\cite{can95}.
In the present calculation, we have 
independently recalculated the three--nucleon
bound state for various realistic nuclear potentials, 
and the details of these calculations are 
illustrated in subsection \ref{3NBS}.

The absorption mechanisms
are specified by the operator
${\cal A}$. To avoid double countings, we avoided 
purely nucleonic intermediate states
in ${\cal A}$. This is because we calculate the final--state 
interactions among the three nucleons using realistic NN potentials,
without performing any kind of subtraction in the nuclear potential.

In this study we consider absorption mechanisms generated by 
pion--nucleon interactions in $p$ waves as described by 
the $\pi$N$\Delta$ and $\pi$NN vertices. 
The nonrelativistic interaction hamiltonians are 

\begin{equation}
H_{\pi N\Delta}= {f_{\pi N\Delta}\over m_\pi}\int \!
d{\bf r} \; \rho({\bf r}) (\vec{S} \cdot \vec{\nabla})
(\vec{\Phi}({\bf r})\cdot \vec{T}) \,,
\end{equation}

\noindent
for the $\pi$N$\Delta$ interaction, and 

\begin{equation}
H_{\pi NN}= {f_{\pi NN}\over m_\pi}\int \!
d{\bf r} \; \rho({\bf r}) (\vec{\sigma} \cdot \vec{\nabla})
(\vec{\Phi}({\bf r})\cdot \vec{\tau})\,,
\end{equation}

\noindent
for the $\pi$NN one. Here,
the baryonic density is denoted by $\rho({\bf r})$,
while $\vec{\Phi}({\bf r})$ is the pionic isovector field.
The quantities $M$ and  $m_\pi$ are the nucleon and
pion  masses, respectively, while $\vec \sigma$ and $\vec \tau$
are the nucleon spin and isospin operators, and 
${\vec S}$ and ${\vec T}$ are the corresponding
generalization to the isobar--nucleon transition.

The $\pi$NN vertex defines the simplest one--body absorption mechanism
and is sometimes referred to as the impulse approximation (IA)
mechanism.
However, this is suppressed
because of energy--momentum mismatch, 
therefore two--nucleon mechanisms dominate.
These are taken into account through the $\Delta$--rescattering
process, where the $\pi$NN--NN inelasticities are modeled
through a $\pi$N$\Delta$--$\Delta$N--NN two-step transition.
Here the first transition is triggered by the $\pi$N$\Delta$
vertex given above, followed by an intermediate $\Delta$N
propagation and by the $\Delta$N--NN transition.
The intermediate propagation of the isobar 
is described by the Green's function

\begin{equation}
G_0={1\over E+M-{\cal M}_\Delta-{p^2\over 2\mu_\Delta}-
{q^2\over 2\nu_\Delta}}\,,
\end{equation}

\noindent
where $E$ is given by the pion energy plus
the target kinetic energy in the c.m. system.
The Jacobi variables $p$ and $q$ are the
pair $\Delta$N and spectator nucleon 
momenta, respectively. Correspondingly,
$\mu_\Delta$ and $\nu_\Delta$ are the reduced masses
of the pair and spectator--pair systems.
${\cal M}_\Delta-M$ is the isobar--nucleon mass gap.
Since the $\Delta$--isobar is not a stable particle
but a resonance, its mass is endowed with an imaginary part
$\Gamma_\Delta$, associated to the decay width of the isobar

\begin{equation}
{\cal M}_\Delta=M_\Delta-E_s-{i\over 2} \Gamma_\Delta(E) \,,
\end{equation}

\noindent
with $M_\Delta$ = 1232 (MeV) and $E_s$ being the energy--shift parameter.
The energy dependence for $\Gamma$
has been modeled phenomenologically
in a previous analysis of the 
$\pi d\leftrightarrow pp$ reaction\cite{can96}.
Here we use the same analytical expression
which has been obtained in that analysis, with the $\Gamma_\Delta$
energy dependence extracted by the condition

\begin{equation}
\sigma(E)={D\over (E-E_R)^2+{\Gamma_\Delta(E)^2\over 4}}\,,
\end{equation}

\noindent
where $\sigma(E)$ is the experimental excitation function
for pion absorption on deuterons. In the numerical calculation,
the fit to $\sigma(E)$ obtained by 
Ritchie \cite{ritchie}
has been used.

In a meson--exchange framework \cite{suga,gari},
the $N\Delta$--$NN$ transition potential can be obtained
from the $\pi$-- and $\rho$--exchange diagrams

\begin{eqnarray}
V_{N\Delta} &=&(V_{N\Delta}^\pi +V_{N\Delta}^\rho)
({{\vec{T_1}}^{\dag}}\cdot\vec{\tau_2}) \nonumber  \\
V_{N\Delta}^\pi &=&-{g_\pi f_{\pi N\Delta}\over 2Mm_\pi}
\, ({\vec{S_1}^{\dag}}\cdot{\vec Q})(\vec{\sigma_2}\cdot{\vec Q})
G(\omega_\pi)
 \nonumber \\
V_{N\Delta}^\rho &=& -{(g_\rho+f_\rho)
f_{\rho N\Delta}\over 2Mm_\rho} \,
({{\vec{S_1}}^{\dag}}\times{\vec Q})\cdot
(\vec{\sigma_2}\times{\vec Q})
G(\omega_\rho)\,.
\end{eqnarray}

The operator ${\vec Q}$ is the baryon--baryon transferred momentum,
$m_\rho$ is the mass of the $\rho$--meson
and the function $G(\omega)$, which describes the meson--exchange
propagation, takes into account the
mass difference between nucleons and isobars

\begin{eqnarray}
G(\omega)&=&
{1\over 2\omega^2}+
{1\over 2\omega(M_\Delta-M+\omega)} \nonumber \\
&\simeq&
{1\over 2\omega^2}+
{1\over 2\omega^2 +2 m_\pi (M_\Delta-M)} \,.
\end{eqnarray}

\noindent
Here, 
$\omega$ is the relativistic energy of the exchanged meson
and, as shown in the last expression, we have taken into account
the $\Delta$N mass difference in an approximated way in order
to obtain analytical expressions when performing the partial--wave
expansion. Each meson--baryon coupling in the transition potential has
been endowed with a phenomenological form factor of monopole type, with
the exclusion of the $\rho N\Delta$ coupling, where a dipole--type
form factor was assumed.

Finally, we discuss how the three--nucleon
dynamics can be incorporated into the theory,
or equivalently, how  we calculate $_S\langle\Psi^{(-)}_{d}|$.
The method is similar to previous procedures developed to incorporate 
final--state interactions (FSI) in the photodisintegration of three--body 
systems \cite{barbour,gibson,sandhas,wolfg,Schadow97a}.
First, we introduce the operator $P_{23}$ which exchanges spin, isospin and 
position coordinates of nucleons 2 and 3. We then introduce the cyclic and
anti--cyclic permutation operators $P_2$ and $P_3$
respectively. They exchange the global coordinates
of the three nucleons in the following manner
$123 \rightarrow 231$ and $123 \rightarrow 312$, and can be expressed
in term of the pair exchange operators as 
$P_2\equiv P_{12} P_{23}$ and $P_3\equiv P_{13} P_{23}$.
This leads to the full permutator

\begin{equation}
\label{permutator}
P\equiv P_2 +P_3\,, 
\end{equation}

\noindent
and the normalized symmetrizer 
$S\equiv {1+P\over \sqrt{3}}$.
It is now possible to derive the complete
wavefunction  in terms of the asymptotic channel wavefunction
${_1\langle}\Phi_{d}|$ (the subscript ``1"
denotes antisymmetrization with respect to the pair made of nucleons
labeled ``2" and ``3")

\begin{eqnarray}
_S\langle\Psi^{(-)}_{d}|= {_1\langle}\Phi_{d}| (1+{\bf T}G_0) S\,,
\end{eqnarray}

\noindent
where the three--body operator ${\bf T}$
satisfies a Faddeev--like equation

\begin{equation}
{\bf T}= P t + P t G_0 {\bf T} 
\end{equation}

\noindent
($t$ is the two--nucleon t--matrix).

\noindent
We now define the M{\o}ller
operator $\Omega^{(-)\dag}\equiv 1+{\bf T}G_0$, in which case

\begin{eqnarray}
_S \langle\Psi^{(-)}_{d}|={_1 \langle}\Phi_{d}| \Omega^{(-)\dag} S \,.
\end{eqnarray}

\noindent
This operator satisfies the Faddeev--like equation

\begin{eqnarray}
\Omega^{(-)\dag}  = 1+PtG_0\Omega^{(-)\dag} \,,
\end{eqnarray}

\noindent
and by its means the pion--disintegration amplitude can be rewritten as

\begin{eqnarray}
A_{d}=
{_1 \langle}\Phi_{d}|\Omega^{(-)\dag} S {\cal A}|\Psi_{\text{BS}}\,,
{\bf P}^{\pi}_0\rangle \,.
\end{eqnarray}

\noindent
Because of the two equations above, 
the full amplitude satisfies the integral equation

\begin{eqnarray}
\label{integraleq}
A_{d}=
{_1 \langle}\Phi_{d}|S {\cal A}|\Psi_{\text{BS}},{\bf P}^{\pi}_0\rangle +
{_1 \langle}\Phi_{d}|PtG_0\Omega^{(-)\dag} S {\cal A}|\Psi_{\text{BS}} \,,
{\bf P}^{\pi}_0\rangle\,.
\end{eqnarray}

\noindent
The first term on the right--hand side
corresponds to the plane--wave contributions, while in the second
term the NN rescatterings (to all orders) are singled out.

As is well known, when the two--body transition 
matrix is represented in a separable form,
the  Alt--Grassberger--Sandhas (AGS) equations 
\cite{AGS} for 
neutron--deuteron scattering reduce to effective
two--body Lippmann--Schwinger equations.
The same happens for the calculation of final--state interaction effects
in the photodisintegration of the triton 
($\gamma +t \rightarrow n+d$)
\cite{barbour,gibson,sandhas,wolfg,Schadow97a},
 where one obtains a similar effective
two--body Lippmann--Schwinger equation, the only difference being that the
driving term of the {\it n-d} scattering equation 
({\it i.e.} the particle--exchange diagram, the so--called ``Z''--diagram) 
is replaced by the off--shell extension of the plane--wave
photoabsorption amplitude.
Here, we use a similar scheme for pion absorption on the 
three--nucleon systems.

In order to accomplish this,
we use the separable expansion method proposed by Ernst, Shakin, and
Thaler (EST) \cite{Erns74} for representing a given NN interaction.
The EST method 
is very reliable and has been tested in the past for bound-state
\cite{parke91,schadow97b}
and scattering calculations \cite{Haid86,Koike87,Corn90}. 
In this approximation the original interaction is expressed
in separable terms of the form

\begin{equation}
  V =  \sum_{\mu,\nu =1}^N |{f_\mu } 
\rangle \, \Lambda_{\mu \nu} 
\, \langle {f_\nu }| \,,
\end{equation}

\noindent
where $N$ is the rank of the approximation. 
The detailed structure of the potential, the parameters for the
formfactors $|f_\mu  \rangle$ and the interaction strength $ \lambda$ can be
found in Ref. \cite{Koike87,Corn90}. 
The input  two--body transition matrix $t$ is given as 
a series of separable terms

\begin{eqnarray}
t=\sum_{\mu \nu}|f_\mu\rangle \,\Delta_{\mu \nu} \, \langle f_\nu| \,.
\end{eqnarray}

To simplify the notation, we restrict the sum above to
just one value for the indices $\mu$ and  $\nu$, so that 
the transition matrix takes the rank--one form 
$t=|f_1\rangle \, \Delta \,\langle f_1|$.
In $\Delta$ as well as in the form factor $|f_1\rangle$, 
we have also omitted
the proper energy dependence,
but it must be remembered that when the two--body $t$--matrix 
is embedded in the three--particle space the correct energy
dependence is upon $E-({3q^2\over 4M})$, where the
energy of the spectator nucleon has to be subtracted.
The separable representation
reproduces the correct negative--energy bound--state pole
of the two--body t--matrix (for the deuteron quantum numbers) 
if the form factor satisfies the homogeneous equation 

\begin{eqnarray}
VG_0(E_d)|f_1\rangle=|f_1\rangle \,,
\label{effeqns}
\end{eqnarray}

\noindent
and, within a normalization factor,
$G_0(E_d)|f_1\rangle$ is the deuteron wave $|\Phi_d\rangle$.

Use of the separable representation of the t--matrix input
in the integral equation (\ref{integraleq})
leads to the effective two--body equation

\begin{eqnarray}
A(q,E)= B(q,E)
+ \int q'^2dq' \,V(q,q',E) \,\, \Delta(E-{3q'^2\over 4M}) \, A(q',E)
\label{quasiparticella}
\end{eqnarray}

\noindent
with the definitions

\begin{eqnarray}
A(q,E) &=&
_1 \langle f_1,q|G_0^{(+)}
\Omega^{(-)\dag} S {\cal A}|\Psi_{\text{BS}},{\bf P}^{\pi}_0\rangle \,, \cr
B(q,E) &=&
_1 \langle f_1,q|G_0^{(+)} S {\cal A}|\Psi_{\text{BS}}\,,
{\bf P}^{\pi}_0\rangle
 \,,
 \cr
V(q,q',E)   &=&
_1 \langle f_1,q|G_0^{(+)}P|f_1,q'\rangle_1 \,.
\end{eqnarray}

\noindent
Here, $A$ and $B$ represent the off--shell
extension for the full and plane--wave pionabsorption amplitudes,
 respectively,
while  $V$ is the effective two--body potential which
represents the one--particle exchange diagram
between different sub--cluster rearrangements.

These amplitudes are decomposed in three--nucleon partial waves,
while the pion--nucleus incident wave is treated in three dimensions.
Details on the employed representation were 
given elsewhere\cite{casveca}. 
The representation of the three--body states is defined in momentum space
and the partial--wave decomposition is discussed
within the $jI$ coupling scheme.
The index $\alpha$ refers to the whole set of quantum numbers
({\it i.e.} orbital momentum, spin, total angular momentum, and isospin 
of the pair $(ls)j;t$, of the spectator $(\lambda\sigma)I;\tau$,
total angular momentum, isospin and associated
third components $JJ^z;TT^z$).

The absorption mechanisms we include in our calculation
have the following structure:

\begin{eqnarray}
\lefteqn{B^\Delta_E(q',\alpha',E)=2\sqrt{3}}& & \nonumber \\
& &\times \sum_{\alpha'',\alpha_\Delta,\alpha}\int 
  p'^2dp' \, p''^2dp'' \, q''^2 dq'' \, p_\Delta^2 dp_\Delta \,
p^2 dp \,q^2 dq \,\,
{f_1(\alpha',p')\, \langle p'q'\alpha'|P|p''q''\alpha''\rangle 
\over E-{3\over 4}{q'^2\over M}-{p'^2\over M}+i\epsilon} \nonumber \\
& & \times \,{
\langle p''q''\alpha''|V_{N\Delta}|p_\Delta q'' \alpha_\Delta\rangle 
\over E+(M-{{\cal M}_\Delta})-{q''2\over 2\nu_\Delta}-
{p_\Delta^2\over 2\mu_\Delta}} \,\,
\langle p_\Delta q''\alpha_\Delta|H_{\pi N \Delta}|
pq\alpha\rangle \, \langle pq\alpha|\Psi_{\text{BS}}\rangle \,,
\end{eqnarray}

\noindent
and

\begin{eqnarray}
\lefteqn{B^\Delta_D(q',\alpha',E)=2\sqrt{3}}& & \nonumber \\
& &\times \sum_{\alpha_\Delta,\alpha}\int
p'^2dp' \,  p_\Delta^2dp_\Delta \, p^2dp \, q^2 dq \,\,
\frac{f_1(\alpha',p')}
{ E-{3\over 4}{q'^2\over M}-{p'^2\over M}+i\epsilon}\nonumber \\
& &\times \, \frac{
\langle p'q'\alpha'|V_{N\Delta}|p_\Delta q' \alpha_\Delta\rangle 
}
{ E+(M-{{\cal M}_\Delta})-{q'2\over 2\nu_\Delta}-
{p_\Delta^2\over 2\mu_\Delta}} \,\,
\langle p_\Delta q'\alpha_\Delta|H_{\pi N \Delta}|
pq\alpha\rangle \, \langle pq\alpha|\Psi_{\text{BS}}\rangle 
\label{deltadirect}
\end{eqnarray}

\noindent
for the $\Delta$--rescattering mechanisms,
and 

\begin{eqnarray}
\lefteqn{B^{I.A.}_E(q',\alpha',E)=\sqrt{3}}& & \nonumber \\
& &\times \sum_{\alpha'',\alpha}\int 
p'^2 dp' \,p''^2 dp''\, q''^2 dq'' \, p^2 dp \, q^2 dq \,\,
\frac{f_1(\alpha',p') \, \langle p'q'\alpha'|P|p''q''\alpha''\rangle }
{E-{3\over 4}{q'^2\over M}-{p'^2\over M}+i\epsilon}
\nonumber \\
& & \times \, \langle p''q''\alpha''|H_{\pi N N}|
pq\alpha \rangle \, \langle pq\alpha|\Psi_{\text{BS}}\rangle \,,
\end{eqnarray}

\begin{eqnarray}
\lefteqn{B^{I.A.}_D(q',\alpha',E) = \sqrt{3}} & &  \nonumber \\
& & \times \sum_{\alpha} \int  p'^2 dp' \, p^2 dp\, q^2 dq \,\,
\frac{f_1(\alpha',p') \,
 \langle p'q'\alpha'|H_{\pi N N}|pq\alpha\rangle \,
\langle pq\alpha|\Psi_{\text{BS}}\rangle }
 { E-{3\over 4}{q'^2\over M}-{p'^2\over M}+i\epsilon} \,,
\end{eqnarray}

\noindent
for the one--body IA mechanisms.
Clearly, all these mechanisms add coherently to give the total amplitude.
The subscripts ``E'' and ``D'' differentiate
between exchange and direct mechanisms, respectively.
The ``D'' mechanisms correspond to the probability
that nucleon ``1'' is the free nucleon in the outgoing channel
(hence nucleons ``2'' and ``3'' form a deuteron) 
while the other two cases ({\it i.e.} nucleon ``2'' or  
``3'' as free outgoing particle) are assembled together in the 
``E'' mechanisms. These exchanges in the rearrangement channel are performed
by the permutation operator $P$. We refer to the appendix of 
Ref. \cite{gloeckle} for details on the partial--wave matrix 
elements for this operator.
Each $\Delta$--rescattering mechanism is multiplied by the factor
$2\sqrt{3}$ which arises from the multiplicity of the possible two--body
mechanisms and by the normalization factor due to Pauli principle.
The same considerations lead to the factor $\sqrt{3}$ in the impulse
approximation. We also observe that
a coupled--channel structure has to be intended for the deuteron form factor
$f_1(\alpha',p)$, since the deuteron $d$--wave component is obviously 
taken into account in our calculation. 
However, these coupled orbital--momentum components
of the pair must be summed up coherently.

We assume that such a sum (over
the coupled $l$'s) is performed at the present stage,
so that the index $\alpha'$, from now on, collectively denotes
the set of quantum numbers $s$, $j$, $t$, $\lambda$, $I$, $\tau$,
$J$, $J^z$, $T$, and $T^z$.

Finally, we observe that the direct contribution to the $\Delta$--rescattering
term, Eq.(\ref{deltadirect}), vanishes on--shell. 
(On-shell, the two--nucleon quantum numbers
$s$, $j$, and $t$ are fixed by the deuteron state, and the relative 
motion of the spectator $q$ is fixed by total energy conservation.)
This vanishing effect is due to isospin considerations, since
the intermediate $\Delta$N pair cannot be directly 
coupled to a final deuteron--like ($t=0$) pair. It does couple, however,
through the permutation operator $P$ in the exchange
contribution, as well as through final--state
interactions, where one can well have isovector pairs in the intermediate
states. 

Once the total absorption amplitude $A^{\text{tot}}(q,\alpha',E)$ 
has been obtained, the pion--absorption excitation function is given by

\begin{equation}
\sigma= {c_{ps} \over 2} \sum_{\alpha'} |A^{\text{tot}}(q,\alpha',E)|^2\,,
\end{equation}

\noindent
with  phase--space factor

\begin{equation}
c_{ps}=(2\pi)^4 {q\over P^0_\pi} {E_\pi E_{t} E_N E_t
\over {E^{\text{tot}}}^2} \,,
\end{equation}

\noindent
where

\begin{equation}
\begin{array}{lcl}
E_\pi  &=& \sqrt{m_\pi^2+{P_\pi^0}^2}\\
E_{t}  &=& \sqrt{M_{T}^2+{P_\pi^0}^2}\\
E_N    &=& \sqrt{M^2+{q}^2}\\
E_d    &=& \sqrt{M_D^2+{q}^2}\\
E^{\text{tot}}&=& E_N+E_d=E_\pi+E_{t}\,.
\end{array}
\end{equation}

\noindent
Here, $M_D$, $M_T$, are the deuteron and three--nucleon masses, 
respectively. The momentum
$q$ is the on--shell momentum (in c.m. frame) of the outgoing nucleon.

Similarly, the unpolarized differential cross--section
for the $\pi^+ t\rightarrow p d$ reaction is  given by

\begin{eqnarray}
\lefteqn{
{d\sigma\over d\Omega}(\theta)
={c_{ps} \over 2} \sum_{K,K^z} \times} 
\nonumber \\
& &|
\sum_{J,J^z,\lambda } C(\lambda K J;\lambda^z K^z J^z) \,
 Y^{\lambda^z}_\lambda (\Omega) \,
 \sum_I (-)^{\lambda-I+K} {\hat K} {\hat I}
\left \{ \begin{array}{ccc}
1        & {1\over 2} & K\\
\lambda  & J          & I\\
       \end{array} \right \}
A^{\text{tot}}(\lambda,I,J,J^z;T,T^z)|^2 \,
\label{diffX}
\end{eqnarray}

\noindent
where K is the channel--spin of the deuteron--nucleon system,
{\it i.e.} the sum of the spins of the two fragments 
${\bf K} = {\bf j} + {\bf \sigma}$.

Finally, detailed balance gives the expression for the unpolarized
cross section for the 
inverse (pion production) reaction, 

\begin{eqnarray}
\left[{d\sigma \over d\Omega}(\theta)\right]^{\text{production}} = 
{1\over 3}
\left({P^0_\pi\over q}\right)^2
\left[{d\sigma \over d\Omega}(\theta)\right]^{\text{absorption}} \,,
\end{eqnarray}
 
\noindent
where $\theta$ is the c.m. angle of the deflected particle.

\subsection{Partial--wave absorption amplitudes}

We outline here the basic ingredients we used for the calculation
of the absorption amplitudes in partial waves.
The section is mainly technical and can be ignored on a first 
reading, if one is not interested in the details of the calculation.

For brevity, out of the four mechanisms discussed previously,
we have selected only two mechanisms, namely the exchange 
$\Delta$--rescattering, which is the dominant one, 
and the direct I.A. term.
The remaining mechanisms, direct $\Delta$--rescattering and 
exchange I.A. have a similar structure and do not introduce any novelty.

To avoid unnecessary complications in the formulas, we give here the 
amplitudes on shell, {\it i.e.}, we have substituted 
$|\Phi_d\rangle$ in place of $G_0(E_d)|f_1\rangle$
and have denoted by $u_{lsj}(p)\equiv u(p)$
the $s$ and $d$ component of the deuteron in momentum space.
For the inclusion of the three--nucleon dynamics via the AGS
equations, these same amplitudes have been extended off shell
(see Eq. (\ref{effeqns})).
To fix the notation, the exchange $\Delta$--rescattering amplitude in 
partial waves can be denoted as follows

\begin{eqnarray}
B_E^\Delta = 
2\sqrt{3}\, \langle u \alpha' q' | P \hat{\cal A}_1|\Psi_{\text{BS}}\,, 
\bar\alpha, {\bf P}^\pi_0\rangle \,,
\end{eqnarray}

\noindent
where $q'$ is the c.m. momentum of the $N$--$d$ system,
${\bf P}^\pi_0$ is the c.m. momentum of the $\pi^+ \,t$ 
channel and  denotes the beam axis or equivalently the z--axis. 
While the former is
a one--dimensional variable in a fully decomposed partial--wave scheme,
the latter is three--dimensional, since we decompose
in partial waves only the baryonic coordinates, not the
pionic one. 
The operator $\hat{\cal A}_1$ represents the $\pi N$--$\Delta$
transition on nucleon labeled ``2", intermediate $\Delta$ propagation,
and a $\Delta N$--$NN$ transition between nucleons ``2" and ``3",
while nucleon ``1" acts as spectator. The process
is diagrammatically illustrated in Figure 1 of Ref.\cite{casveca},
while $P$ is the three--body permutator previously introduced.
For the detailed form of $P$ in partial waves, we used the expression
given in the appendix of Ref.\cite{gloeckle}.

With $\alpha'$ we collectively denote the quantum numbers for
the $jI$ decomposition of the $p$--$d$ channel listed according to
Eq. (3.2) of ref.\cite{casveca}, while
with $\bar\alpha$ the quantum numbers for the three--nucleon
bound state  in the $LS$ scheme are assumed. The details
of the calculation in this scheme can be found in
Ref.\cite{sauer}, while similar calculations in the $jI$ scheme
are found {\it e.g.} in Refs.\cite{sasa,gloeklebook}.

The resulting expression is

\begin{eqnarray}
\lefteqn{
\langle u, \alpha', q' |
 P \hat{\cal A}_1|\Psi_{\text{BS}}, \bar\alpha, {\bf P}^\pi_0\rangle = 
{\cal F} 
\sum^{\ }
_{\stackrel{k l_1 l_1' \alpha'' \alpha_\Delta}
  { L_\Delta S_\Delta S_z m}}
\tilde g(\alpha' \alpha'' k l_1 l_2 l_1' l_2') \,\,
{\cal T}(\alpha'' \bar\alpha) 
} &&
\nonumber \\
& & \times \,
{\cal B}(\alpha'' \alpha_\Delta \bar\alpha L_\Delta S_\Delta S_z m) \,\,
{\cal I}_1(\alpha' \alpha'' \alpha_\Delta 
\bar\alpha k l_1 l_1' L_\Delta S_\Delta S_z m) \,, 
\end{eqnarray}

\noindent
where

\begin{eqnarray}
{\cal F}={i P_0^\pi (3M+m_\pi)f_{\pi N\Delta}^2 f_{\pi N N}
          \over
          72 \pi^3 m_\pi^3 (M+m_\pi) \sqrt{\pi \omega_\pi} }\,,
\end{eqnarray}

\begin{eqnarray}
\tilde g&&
(\alpha \alpha' k l_1 l_2 l_1' l_2')= 
 - \hat l\hat s\hat j \hat t\hat \lambda\hat I
\hat l'\hat s'\hat j' \hat t'\hat \lambda'\hat I'
\left\lbrace
\matrix
{1\over 2& 1\over 2& t\cr
 1\over 2& T       & t'}
\right\rbrace
 \nonumber
\\
&
\lefteqn{\times \,
\sum_{LS}(-)^L {\hat L}^2{\hat S}^2
\left\lbrace
\matrix
{1\over 2& 1\over 2& s\cr
 1\over 2& S       & s'}
\right\rbrace
\left\lbrace
\matrix
{ l & s & j \cr
 \lambda & 1\over 2 & I \cr
  L & S & J }
\right\rbrace
\left\lbrace
\matrix
{ l' & s' & j' \cr
 \lambda' & 1\over 2 & I' \cr
  L & S & J }
\right\rbrace
\nonumber
}& \ 
\\
&
\lefteqn{ \times \,
(-)^{k+\lambda'+l_1'} \,
({\textstyle{1\over 2}})^{l_1+l_2'}\,
({\textstyle{3\over 4}})^{l_2}\,
\sqrt{(2l'+1)!
       \over
      (2l_1)!(2l_2)!}
\sqrt{(2\lambda'+1)!
       \over
      (2l_1')!(2l_2')!}
\nonumber
} & \ 
\\
&
\lefteqn{\times \,
\sum_{ff'}\hat f \hat f' \,
C(l_1 l_1' f;000) \,
C(l f k;000) \,
C(l_2 l_2' f';000) \,
C(\lambda f' k;000)
\nonumber
} \ 
\\
& 
\lefteqn{\times \,
\left\lbrace
\matrix
{ l  & f & k \cr
  f' & \lambda & L }
\right\rbrace
\left\lbrace
\matrix
{ l_1 & l_2 & l' \cr
  l_1' & l_2'& \lambda' \cr
  f &  f' & L }
\right\rbrace
\ , }& \ 
\label{pistolotto}
\end{eqnarray}

\begin{eqnarray}
{\cal T}&& 
(\alpha' \bar\alpha)=
96 \, \sqrt{30} \, (-)^{s'+j'+1+\bar s+\bar t}
\hat s' \hat l' \hat j' \hat t' \hat I' 
\hat{\bar s} \hat{\bar S} \hat{\bar t} \hat{\bar T} 
\nonumber\\
&\lefteqn{\times \,
\left\lbrace
\matrix{t' &1\over 2&1\over 2\cr
        1  &3\over 2&1\over 2}
\right\rbrace
\left\lbrace
\matrix{1\over 2& \bar t & t' \cr
        1\over 2&T'&\bar T}
\right\rbrace
\left\lbrace
\matrix{1 &1\over 2&3\over 2\cr
        1\over 2&t'&\bar t}
\right\rbrace
C(1\bar T T'; 1 {\textstyle-{1\over 2}}  {\textstyle{1\over 2}})
\ ,}
\end{eqnarray}

\begin{eqnarray}
{\cal B}&&(\alpha' \alpha_\Delta \bar\alpha L_\Delta S_\Delta S_z m)=
(-)^{s_\Delta+S_\Delta+\bar T} i^{l_\Delta-l'}
\hat{s}_\Delta^2\hat{l}_\Delta\hat{S}_\Delta\hat{L}_\Delta
\left (
\matrix{ l' & 2 & l_\Delta \cr
         0  & 0 & 0 }
\right )
\left\lbrace
\matrix{1 &1\over 2& 3\over 2\cr
        1\over 2 &s_\Delta&\bar s}
\right\rbrace
\nonumber
\\
&\lefteqn{\times \,
\left\lbrace
\matrix{1 &\bar s& s_\Delta\cr
        1\over 2 &S_\Delta&\bar S}
\right\rbrace
\left\lbrace
\matrix{j' & s' & l' \cr
        2 &l_\Delta&s_\Delta}
\right\rbrace
\left\lbrace
\matrix{1\over 2& 3\over 2&1\cr
        1\over 2&1\over 2 &1\cr
        s'      &s_\Delta &2}
\right\rbrace
\left\lbrace
\matrix{l_\Delta &s_\Delta &j' \cr
        \lambda' &1\over 2 &I' \cr
        L_\Delta &S_\Delta &J' }
\right\rbrace
}
\nonumber
\\
&\lefteqn{\times \,
C\Big(L_\Delta S_\Delta J';(J_z'-S_z)S_zJ_z'\Big) \,
C\Big(\bar L\bar S\bar J;(J_z'-S_z)S_zJ_z'\Big) \,
C\Big(1 \bar S S_\Delta ;0 S_z S_z\Big)
}\nonumber\\
&
\lefteqn{ \times \,
C\Big(l_\Delta \lambda' L_\Delta ;m (J_z'-S_z-m)(J_z'-S_z)\Big) \,
C\Big(\bar l \bar \lambda \bar L;m (J_z'-S_z-m)(J_z'-S_z)\Big)
\ ,}
\end{eqnarray}

\noindent
and         

\begin{eqnarray}
{\cal I}_1(\alpha' \alpha'' \alpha_\Delta 
\bar\alpha k l_1 l_1' && L_\Delta S_\Delta S_z m)=
{q'}^{l_2+l'_2}
\int\limits_0^\infty {p'}^{2} dp' \, {p'}^{l_1+l_1'} \, u_{l's'j'}(p')
\int\limits_{-1}^{1} dx \, {P_k(x)\over {p^*}^{l''} {q^*}^{\lambda''}}
\int\limits_0^\infty p^2 dp
\nonumber\\
&
\lefteqn{\times\,
\int\limits_{-1}^{1} d\cos{\hat {\bf P}} \,
{{\cal Q}^T_{l''l_\Delta}(p^*,p_\Delta) 
\over 
  E-{p_\Delta^2\over 2\mu_\Delta}-{{q^*}^2\over 2\nu_\Delta}} \,
\Theta_{l_\Delta m}(\cos{\hat {\bf P}_\Delta}) \,
\Theta_{\bar l m}(\cos{\hat {\bf P}})
}
\nonumber\\
&
\lefteqn{\times \,
\int\limits_{-1}^{1} d\cos{\hat {\bf Q'}} \,
\Theta_{\lambda'' n}(\cos{\hat {\bf Q'}}) \,
\Theta_{\bar \lambda n}(\cos{\hat {\bf Q}}) \,
\Psi_{\bar \alpha}(p,q)
\ .}
\label{bigint1}
\end{eqnarray}

In this last equation,
all the relevant momenta are defined 
in terms of the channel momenta $q'$ and $P_0^\pi$
and of the integration variables
through the formulas

\begin{eqnarray}
p^*&=&\sqrt{{\textstyle{1\over 4}} {p'}^2+{\textstyle{9\over 16}}{q'}^2 + 
{\textstyle{3\over 4}} p'q'x} \,,\nonumber\\
q^*&=&\sqrt{ {p'}^2+{\textstyle{1\over 4}}{q'}^2-p'q'x}\,, \nonumber \\
p_\Delta
&=&
\sqrt{({3M+m_\pi\over 6M+3m_\pi}{ P}^\pi_0)^2
+p^2+2{3M+m_\pi\over 6M+3m_\pi}{ P}^\pi_0 \,p \cos{\hat{\bf P}}}\,,
\nonumber \\
\cos{\hat{\bf P}}_\Delta
&=&
{{3M+m_\pi\over 6M+3m_\pi}{ P}_0^\pi+p \cos{\hat{\bf P}}\over
p_\Delta}\,,
\nonumber \\
q
&=&
\sqrt{({\textstyle{1\over 3}}{ P}^\pi_0)^2
+{q^*}^2+{\textstyle{2\over 3}}{ P}^\pi_0 \, {q^*}^2 \cos{\hat{\bf Q'}}}\,,
\nonumber \\
\cos{\hat{\bf Q}}
&=&
{{1\over 3}{ P}_0^\pi+q^* \cos{\hat{\bf Q'}}\over
q}\,.
\end{eqnarray}

In Eq. (\ref{bigint1}) $P_k(x)$ represents the Legendre polynomials,
$\Theta_{lm}(\cos{\theta})$ denotes the 
associated Legendre functions (normalized according to
Ref. \cite{casveca}),
${\cal Q}^T_{l l'}(p,p')$ represents linear combinations
of II--kind Legendre polynomials (see Eq. (43) in Ref. \cite{can96}),
originated by the $\pi + \rho$ meson--exchange 
diagrams in the tensor $\Delta$N--NN force, 
and finally $\Psi_{\bar\alpha}(p,q)$
is the triton wavefunction in momentum space
and in the $LS$ scheme. 
In the same equation, it is also assumed that
$n=J'_z-S_z-m$, while
the identities $l_1+l_2=l'$ and 
$l_1'+l_2'=\lambda'$ hold in both
Eqs. (\ref{pistolotto}) and (\ref{bigint1}).

Similarly, the direct I.A. amplitude in partial waves
can be expressed as follows

\begin{eqnarray}
B_D^{I.A.} = 
\sqrt{3}\, \langle u \alpha' q' | H_{\pi NN}|\Psi_{\text{BS}}, 
\bar\alpha, {\bf P}^\pi_0\rangle \,,
\end{eqnarray}

\noindent
with the $\pi$NN vertex acting on nucleon ``2" as a
one--body operator.

The calculation of this matrix element yields

\begin{equation}
\langle u, \alpha', q' |
 H_{\pi NN} |\Psi_{\text{BS}}, \bar\alpha, {\bf P}^\pi_0\rangle =
{\cal F}_2 \,
\sum^{\ }_{L' S' S_z m}
{\cal B}_2(\alpha' \bar\alpha L' S' S_z m) \,\,
{\cal I}_2(\alpha' \bar\alpha L' S' S_z m) \,, 
\end{equation}

\noindent
where

\begin{eqnarray}
{\cal F}_2={i P_0^\pi (3M+m_\pi) f_{\pi N N}
          \over
          2 \pi m_\pi (M+m_\pi) \sqrt{\pi \omega_\pi} } \,,
\end{eqnarray}

\begin{eqnarray}
\lefteqn{
{\cal B}_2 (\alpha' \bar\alpha L' S' S_z m)  =
(-)^{1+\bar s+s'+\bar t+t'+S'+T'} 
\hat{\bar s}\ \hat s'\ \hat{\bar t}\ 
\hat  t'\
\hat{\bar S}\  \hat{\bar T}\ 
\hat j'\ \hat I'\ \hat L'\ 
\hat S' } &&   \nonumber\\
& &{\times \,
\left\lbrace
\matrix{1\over 2& \bar t & t' \cr
        1\over 2&T'&\bar T}
\right\rbrace
\left\lbrace
\matrix{1 &1\over 2&1\over 2\cr
        1\over 2&t'&\bar t}
\right\rbrace
C(1\bar T T'; 1 {\textstyle{-{1\over 2}}} {\textstyle{1\over 2}})
} \nonumber\\
& &{\times \,
\left\lbrace
\matrix{1\over 2& \bar s & s' \cr
        1\over 2&S'&\bar S}
\right\rbrace
\left\lbrace
\matrix{1 &1\over 2&1\over 2\cr
        1\over 2&s'&\bar s}
\right\rbrace
\left\lbrace
\matrix{l' &s' &j' \cr
        \lambda' &1\over 2 &I' \cr
        L' &S' &J' }
\right\rbrace
} \nonumber\\
& &{\times \,
C\Big (L' S' J';(J_z'-S_z)S_zJ_z'\Big) \,
C\Big(\bar L\bar S\bar J;(J_z'-S_z)S_zJ_z'\Big) \,
C\Big(1 \bar S S' ;0 S_z S_z\Big)
}\nonumber\\
& &
{\times \,
C\Big(l' \lambda' L' ;m (J_z'-S_z-m) (J_z'-S_z)\Big) \,
C\Big(\bar l \bar \lambda \bar L;m (J_z'-S_z-m)(J_z'-S_z)\Big)} 
\end{eqnarray}

\noindent
and         

\begin{eqnarray}
{\cal I}_2(\alpha' \bar\alpha && L' S' S_z m)=
\int\limits_0^\infty {p'}^{2} dp' \, u_{l's'j'}(p')
\int\limits_{-1}^{1} d\cos{\hat {\bf P'}} \,\,
\Theta_{l' m}(\cos{\hat {\bf P'}}) \,
\Theta_{\bar l m}(\cos{\hat {\bf P}}) 
 \nonumber\\
&\lefteqn{\times\,
\int\limits_{-1}^{1} d\cos{\hat {\bf Q'}}\,\,
\Theta_{\lambda' n}(\cos{\hat {\bf Q'}}) \,
\Theta_{\bar \lambda n}(\cos{\hat {\bf Q}}) \,
\Psi_{\bar \alpha}(p,q)
\, .}
\label{bigint2}
\end{eqnarray}

\noindent
As in the previous case, all the relevant momenta have to be expressed
in terms of the channel momenta $q'$ and $P_0^\pi$ and of the
integration variables

\begin{eqnarray}
p&=&\sqrt{{p'}^2+
({3M+m_\pi\over 6M+3m_\pi}{{ P}_0^\pi})^2-
2{3M+m_\pi\over 6M+3m_\pi}{ P}^\pi_0 \,p' \cos{\hat{\bf P}'}} \,,
\nonumber\\
q
&=&
\sqrt{{q'}^2 + ({\textstyle{1\over 3}}{ P}^\pi_0)^2
+{\textstyle{2\over 3}}{ P}^\pi_0 \, {q'}^2 \cos{\hat{\bf Q'}}} \,,
\nonumber\\
\cos{\hat{\bf P}}
&=&
{-{3M+m_\pi\over 6M+3m_\pi}{ P}_0^\pi+p' \cos{\hat{\bf P}'}\over
p} \,,
\nonumber\\
\cos{\hat{\bf Q}}
&=&
{{1\over 3}{ P}_0^\pi+q' \cos{\hat{\bf Q}'}\over
q} \,.
\end{eqnarray}

\subsection{3N Bound--state calculation}
\label{3NBS}

For the calculation of the three-body bound-state energy $E_T$
and wavefunction $|{\Psi_{\text{BS}}} \rangle$ 
we start from the Lippmann-Schwinger equation

\begin{equation}
 |{\Psi_{\text{BS}}}\rangle = \lim_{\epsilon \to 0} \,
 i\epsilon \,G(E_T + i\epsilon) |{\Psi_{\text{BS}}} \rangle\;.
\end{equation}

\noindent
Using the resolvent equation

\begin{equation}
  G(z) = G_\beta(z) + G_\beta(z)\, \bar V_\beta\, G(z)
\end{equation}

\noindent
and performing the $\epsilon$-limit we end up with

\begin{equation}
 |{\Psi_{\text{BS}}}\rangle = G_{\beta}(E_T) (V - V_{\beta})|
{\Psi_{\text{BS}}} \rangle \;,
\end{equation}

\noindent
where $V$ is the total interaction summed over the pairs,
$V_\beta$ is the nuclear interaction of the pair labeled
by $\beta$, $\bar V_\beta$ denotes $V - V_\beta$, 
and $G_\beta$ is the channel resolvent
for the two--cluster partition.

If we now introduce the position 
 $ |{F_\beta}\rangle =  (V - V_\beta) \; |{\Psi_{\text{BS}}}\rangle$ 
and the relation $V_\gamma G_\gamma = T_\gamma G_0$, we obtain 
the celebrated equation

\begin{equation}
 |{F_\beta}\rangle = \sum_\gamma (1 - \delta_{\beta \gamma})\,
 T_\gamma \,G_0\,| {F_\gamma} \rangle \,.
\label{celebrated}
\end{equation}

\noindent
Here, the summation runs over all two--fragment
partitions $\gamma$.
The ``form-factors'' $|{F_\beta}\rangle$ are related to $|{\Psi_{\text{BS}}}
\rangle$
by

\begin{equation}
 |{\Psi_{\text{BS}}}\rangle  = \sum_\gamma G_0 \, T_\gamma \, G_0 \, |{F_\gamma} \rangle
= \sum_\gamma  |{\Psi_\gamma}\rangle \,,
\label{psi}
\end{equation}

\noindent
where the $|{\Psi_\gamma}\rangle  $ are the standard Faddeev components. This
can be shown by using the definition of  $|{F_\beta}\rangle$ and
again the relation $V_\gamma G_\gamma = T_\gamma G_0$:

\begin{equation}
 |{\Psi_\gamma}\rangle  =  G_0 \, T_\gamma \, G_0 \, |{F_\gamma} \rangle
= G_0 \, V_\gamma \, G_\gamma \, |{F_\gamma} \rangle 
= G_0 \, V_\gamma G_\gamma \, (V - V_\gamma) | \Psi_{\text{BS}} \rangle 
= G_0 \, V_\gamma  | \Psi_{\text{BS}} \rangle \,.
\end{equation}

\noindent
The bound--state equation (\ref{celebrated}) has to be projected onto
the three--body partial waves $| p q l b J T \rangle$.  
In this subsection we use the so called channel--spin coupling.
The label $b$ 
denotes the set of quantum numbers $(\eta K \lambda)$ , 
where $K$ and $\lambda$ are the
channel spin of the three nucleons [ with the coupling sequence 
$(j,\sigma) K$ ] and the relative angular momentum between the two--body
subsystem and the third particle, respectively. In this paper,
the channel--spin coupling $K$ has been already introduced in 
Eq. (\ref{diffX}). Here, 
the total angular momentum of the three--nucleon system $J$ 
is given by the coupling sequence $(K,\lambda)$ 
while $T$ has been already defined
as the total isospin. $l$ is the usual orbital angular momentum, and 
$\eta = (s,j;t)$ collectively denotes the spin, 
angular momentum [ with the coupling sequence $(l,s)j$], and isospin 
of the two--body subsystem.
Defining

\begin{equation}
 F^{b}_\beta (q) = \sum_l \int\limits^\infty_0  dp \, p^2 \,
f_{l} (p) \, \langle p q l b J T | G_0 | F_\beta \rangle \,,
\end{equation}

\noindent
inserting the finite--rank form for the T-matrix,
and using the Pauli principle, we obtain the 
one--dimensional equation

\begin{equation}
  F^{b} (q) = \sum_{ b'} \int \limits^\infty_0 dq' \, q'^2 \,
 {\cal A V}^{ b b'} (q, q', E_T) \, \Delta^{\eta'}
 (E_T - {\textstyle \frac{3}{4}} q'^2)  \, F^{ b'} (q') \,,
\label{form}
\end{equation}

\noindent
with

\begin{equation}
 {\cal A V}^{b b'} (q, q', E_T) = 2
\sum_{l l'} \int \limits^\infty_0 \!
 \int \limits^\infty_0  dp \, p^2 \,  dp' \, p'^2
\, f_l(p)
\, \langle p q l b J T | G_0 | p' q' l' b' J T \rangle
\, f_{l'}(p') \,.
\label{effpot}
\end{equation}

\noindent
The recoupling coefficients for the coupling scheme herein
employed can be directly found in Ref. \cite{Janu93},
or alternatively, they can be obtained 
starting from other coupling schemes \cite{gloeckle} 
by means of the usual transformation algebra.

Equation (\ref{form}) can be treated as an eigenvalue problem. The energy 
is varied until the corresponding eigenvalue is 1.
The ranks for the partial waves taken into
account for the bound state calculations for the different potentials
are listed in Table \ref{ranktab}.
Results for the binding energies can be found in Table \ref{energytab}.

The whole wavefunction can now be calculated by either using Equation
(\ref{psi}) or by applying the
permutation operator $P$ defined in Eq. (\ref{permutator}) 
on one Faddeev component \cite{gloeckle}

\begin{equation}
| \Psi_{\text{BS}} \rangle = (1 + P) | \Psi_1 \rangle \,.
\end{equation}

\noindent
This second representation has the advantage
to be explicitely independent from the T--matrix 
representation, and is therefore computationally
more convenient when the rank of the representation
becomes large.
The high accuracy of the 
wavefunction has been shown in Refs. \cite{parke91,schadow97b}.

For the calculation of the Bornterms as illustrated in 
the previous section, the wavefunction
was transformed into the $LS$ scheme via 
\begin{eqnarray}
& & \langle {((l\lambda) L, (s \sigma) S) J J^z }|
\nonumber  \\
& = & \sum_{j K} (-1)^{l + s + \lambda + \sigma 
     + L + S + 2 K}
\hat j \hat L \hat S \hat K
 \left\{ \!\!\begin{array}{ccc} {l}&{{S}}&{K} \\
   {\sigma}&{j}&{s} \end{array} \!\!\right \}
 \left\{ \!\!\begin{array}{ccc} {l}&{{S}}&{K} \\
  {J}&{\lambda}&{{L}}  \end{array} \!\!\right \}
\langle {(((l s) j \sigma) K \lambda) J J^z}| \,. \nonumber \\
\end{eqnarray}

\noindent

\subsection{3N continuum calculation}

In this section we briefly summarize the continuum calculation.
As already mentioned the method used here is similar to the
techniques developed to incorporate  final--state interactions (FSI)
in the photodisintegration of triton
 \cite{barbour,gibson,sandhas,wolfg,Schadow97a}.
The only difference is the inhomogeneity, {\it i.e.}, the Bornterm,
of Eq. (\ref{quasiparticella}). Thus, with this replacement our working
program for photodisintegration has been easily modified 
for pionabsorption. Both programs are based on a program
for  $n$--$d$ scattering \cite{Janu93}, which has been rewritten
for the EST-method used here and the above replacement.
The Faddeev results for the continuum 
presented here are all calculated using a rank--one
representation of the two-body t-matrices in the kernel of the
integral equation. In the case of photodisintegration it has been
shown \cite{schadow97c} that this is enough for the observables
calculated.

In the case of the pionabsorption the Bornterm is much more complicated
and has more structure than in the case of photodisintegration.
To achieve the required accuracy we used 70 grid points in the $q$ variable
for each channel for the off-shell extension of the Bornterm.
For details of the numerical solution of the integral equation
we refer to Ref. \cite{Janu90}.

\section{Results}

Our analysis of the $pd\rightarrow\pi^+ t$ reaction is
essentially a {\em parameter free} analysis,
because all the parameters of the model
were fixed in previous studies 
\cite{can96,dor97}
on the simpler $\pi^+ d\rightarrow pp$ reaction.
This was possible because the character of the calculation 
is sufficiently microscopic
that the tuning parameters are basic quantities
such as coupling constants and cut-offs at the meson--baryon vertices.
Obviously these very same parameters enter in both 
$pp\leftrightarrow \pi^+ d$ and $pd\leftrightarrow\pi^+ t$
reactions. We refer to the analyses of Refs.\cite{can96,dor97}
for a more detailed discussion
on the $\pi^+ d\rightarrow pp$ reaction; here we limit ourselves
to summarize few aspects which are specific of this approach.
The coupling constants for the $\Delta$--rescattering
mechanism are the ones referring to the
$\pi NN$, $\pi N \Delta$, $\rho NN$ and $\rho N\Delta$
vertices, which were taken from Table B.1
of Ref.\cite{mac89} (model III).
The cut--off parameters (in MeV) of the four vertices
are 1600, 900, 1200, and 1350, respectively,
which are practically equivalent to the ones given
in the same Table B.1 cited above.

In the analysis of pion absorption on deuterons which we used as input
for our study, there were only two authentically 
phenomenological ingredients.
First, the way the $\Delta$--resonance 
propagates in the intermediate states, and,
second, the off-shell nature of the 
$\pi NN$ and $\pi N\Delta$ vertices
when the pion is the asymptotic particle, 
and hence rigorously standing on its mass shell.

Concerning the first point, the resonance peak
in the experimental pion--deuteron absorption cross--section is 
below the position of a free $\Delta$ (as observed by $\pi N$ scattering)
by a few tenths of MeV.
This effect could not be explained within
the dynamical model, and therefore the $\Delta$ resonance peak has been
positioned downwards by introducing an energy shift parameter.
Phenomenologically, 
the magnitude of this shift is of the order of 30-35 MeV;
it has, however, to be increased of about 20 MeV, 
(see  Refs.\cite{can96,dor97}) if the motion of the 
other nucleon is taken into account.
As for the isobar width, this was fixed to 115 MeV at
the resonance, and therefore corresponds to the value of the free
resonance width. Note, however, that the resonance peak
for pion absorption on deuteron is broader, being around $\Gamma_0=150$ 
MeV \cite{eri88,rit83}. This broadening of
the width is fully described by the $\pi d \rightarrow pp$
model calculation, once all the mechanisms, including FSI,
are taken into account.

The second point with a certain deal of phenomenology 
concerns the off--shell nature of the pion--baryon vertices, in case the 
pion is the asymptotic particle. When the pion is on its mass shell,
the extended structure of the $\pi NN$ and $\pi N\Delta$
vertices may be governed by the nucleon momentum, if the nucleon is the 
off--shell particle. The importance of including these baryonic off--shell
effects has been pointed out also in a recent study of $\pi N$ scattering
\cite{sch94}.
Thus, the two vertices have been endowed 
with a form factor depending upon the momentum ${\bf k}$ 
of the interacting nucleon, 

\begin{equation}
F_{N,\Delta} (k) = {\lambda_{N,\Delta}^2 \over 
\lambda_{N,\Delta}^2 + k^2} \,.
\end{equation}

For the two--nucleon system $k^2$ corresponds to the square of the
Jacobi coordinate ${\bf p}$, while for the three--nucleon
system, an averaging over angles yields
$k^2\simeq p^2+0.25q^2$.

At the level of the two--nucleon system,
the model has been tested against the
enormous variety of experimental data collected for the
$\pi^+ d\leftrightarrow pp$ reaction,
and its strength and weakness, especially at the level
of spin observables, 
have been thoroughly discussed in Refs. \cite{can96,dor97}.

To exhibit the quality of the results at the level
of the two--nucleon system,
we report in Figure 1 the total production cross section
for the $pp\rightarrow \pi^+ d$ reaction from threshold to
above the $\Delta$ resonance. The parameter $\eta$
corresponds to the pion momentum (c.m.), in units of pion mass.
The experimental data were extracted from a collection of 
Refs. \cite{hu91,he96,ro67,ri93,ri81,ri83,go93,ax76,ri70,bo82}.
All data have been converted into production data using the principle of 
detailed balance, if necessary, and in case of the data
of Ref.\cite{hu91}, charge--independence considerations were applied.
The full line is the result obtained with the Bonn {\it B} potential,
and with the $p$--wave pion--nucleon mechanisms which, as
discussed in the previous section, include both isobar and non--isobar 
degrees of freedom. Practically indistinguishable from the full line
there is a dotted line corresponding to a similar calculation but with 
the Paris potential. The differences between the two curves 
slightly increase with energy and at the resonance peak the Paris 
result is less than 5\% smaller than the Bonn {\it B} one. 

The other two curves (dashed and dashed dotted, respectively) 
correspond to the additional inclusion of $s$--wave
pion--nucleon interactions
in the final state of the production process
(or conversely in the initial state, for pion absorption). 
The leading contribution for this $\pi N$--$\pi N$ interaction was obtained
via a $\rho$--meson exchange model in Ref.\cite{dor97}.
The two curves differ in the choice of the NN potential, the Bonn {\it B} 
being referred to by the dashed line, the Paris by
the dotted--dashed curve.

The log scale emphasizes the threshold results, and 
shows that in order to achieve a correct reproduction
of data over three orders of magnitude, the pion--nucleon
interactions via $\rho$--meson exchange  
have to be considered as well.
Note, however, that the process around the $\Delta$ resonance is
dominated by the $p$--wave mechanisms which are triggered by the
$\pi NN$ and $\pi N\Delta$ vertices. On a linear scale,
the threshold effects in pion production due to the $s$--wave mechanisms
would appear smaller.

At this stage of our study of the $pd\leftrightarrow\pi^+ t$
process, we have excluded the $\rho$--mediated pion--nucleon
interaction in $s$--wave, and limited the analysis
around the $\Delta$ resonance. Further work is needed to include
the low--energy mechanisms, and to extend the calculation at 
the production threshold. 
In what follows we considered pion--nucleon
interactions in $p$--waves only. However, both 
$\Delta$ and non--$\Delta$ degrees of freedom 
were considered.
We note, however, that we treat the pion--nucleus
wave three--dimensionally and carefully calculate the relevant kinematical
transformations from the pion--nucleus to the pion--nucleon variables.
This requires additional integrations over angular variables
in order to produce the absorption amplitudes
(see Eqs. (\ref{bigint1}) and (\ref{bigint2})). And,
because of this, the total angular momentum $J$ is not restricted, 
neither is the parity $P$, 
and thus we have taken into account
a few $J^P$ states: 
$J=({1\over 2})^\pm$,
$J=({3\over 2})^\pm$, 
$J=({5\over 2})^\pm$, and  
$J=({7\over 2})^\pm$.
For each of these three--body quantum numbers,
one should consider in principle an infinite set of partial waves
for the two--body subsystem, which matches a corresponding infinite set
of partial waves for the spectator particle to give a fixed $J^P$.
Clearly, a truncation over a limited number of states is necessary.
Table 1 shows the smallest set of three--body partial waves which have been 
included in our calculation. Here, four two--nucleon states
have been included, namely $^1S_0$, $^3S_1$, $^3D_1$, and  $^1D_2$,
while the maximum values considered for the 
orbital and angular momentum of the {\em spectator} 
were $\lambda=5$ and $I={9\over 2}$, respectively,
for a total amount of 82 states. Such three-body states are
indicated in the table according to the notation
$^{2 s+1}{\large l}_j\ {\large \lambda}_I$.

All our calculations, unless otherwise explicitly indicated,
have been performed with a much larger set. In particular,
in calculating the matrix elements of the absorption/production
matrix elements with exchange operator, 
we took into account 18 two--body partial
waves 
$^1S_0$, $^3S_1$, $^3D_1$, $^1D_2$, 
$^3P_0$, $^1P_1$, $^3P_1$, $^3D_2$, 
$^3P_2$, $^3F_2$, $^1F_3$, $^3D_3$, $^3G_3$, 
$^3F_3$, $^3G_4$, $^1G_4$, $^3F_4$, and $^3H_4$.
This, with the same cuts in the spectator quantum numbers $\lambda$ and $I$
specified as before,
yields
34 channels for $J^P=({1\over 2}^\pm)$,
58 for $J^P=({3\over 2}^\pm)$, 
70 for $J^P=({5\over 2}^\pm)$, and  
70 for $J^P=({7\over 2}^\pm)$
for a grand total of 464 three--body states.
In solving the Faddeev 
equations for the three--nucleon dynamics in the $p$--$d$ channel,
of the 18 nucleon--nucleon waves we have included
the first 10 states  $(j \leq 2)$ since
for these the NN state--dependent interaction was available.

Figure 2 shows the integral cross section (in $\mu$b) 
of the $p d\rightarrow \pi^+ t$
reaction around the resonance. The $x$--axis dependence
is upon the parameter $\eta$, which has been previously defined.
The experimental bars were obtained with the help
of a collection of data contained in Refs.\cite{machner1,machner2}.
To this collection, we have added the experimental results
of Ref.\cite{weber89} for $\pi^-$ absorption on $^3{\text{He}}$ at 64 and 119
MeV, assuming time reversal and charge symmetry.
The full, dotted, and dashed lines differ among each other
for the NN potential which has been used to generate
the asymptotic bound states, {\it i.e.} the deuteron and triton. 
The three curves represent
calculations with Bonn {\it B} (solid line), Bonn {\it A} (dotted),
and Paris interactions. We find that
at the peak the Paris results are smaller
than the Bonn {\it B} results by almost 25\%.
This relative difference is substantially larger than
the 5\% difference obtained for
the simpler $pp \rightarrow \pi^+ d$ reaction. Experimentally,
the $p d \rightarrow \pi^+ t$ cross--section 
is smaller by a factor 80$-$85 with respect to
the $p p \rightarrow \pi^+ d$ one.
This large suppression, due to the small overlap
of the deuteron wavefunction
in the incoming channel 
with the pion production matrix elements,
is fully reproduced by our calculations.

The unpolarized differential cross section
is reported in Figure 3 for a variety of energies
spanning the $\Delta$ resonance. 
While Figure 2 shows the normalization of the cross section,
in Figure 3 we have addressed our attention to the pure angular
dependence and therefore all the curves are normalized
to the experimental data.
Thus, the three curves referring to 300 MeV 
(lab energy of the proton) have been multiplied by a factor.
The full line, referring to the result with Bonn {\it B} potential,
has been multiplied by 1.075, the dotted line (Bonn {\it A})
by 1.016, and the dashed line (Paris) by 1.329.
The three factors are due to the differences
in normalization immediately perceived in Figure 2
for $\eta=1$, which is the corresponding value for that energy.

At the resonance peak, the angular differences between the various
NN potentials are practically zero.
These differences, however, increase in moving away from the resonance
in both directions. We find larger differences 
above the resonance, for backward angles.
In particular, calculations with Paris potential seem to reproduce
better the data at backward angles.

In the first sector (E = 300 MeV) of the figure, the data
were taken from the inverse
absorption reaction, $\pi^-$ $^3\text{He}\rightarrow nd$
at 64 MeV,
Ref.\cite{weber89},
using the detailed balance
and charge symmetry.
In the same picture we have reported 
also four points obtained from Ref. 
\cite{lol82}
for the 
production reaction at 305 MeV. The four points are 
clearly visible because they stand above the rest of the data set.
 From the same reference \cite{lol82},
we have taken  also the data at 330 MeV, shown in the second sector,
while the data at 382 MeV were obtained 
from pion absorption reactions at 119 MeV. In particular,
data from both reactions $\pi^-$ $^3{\text{He}}\rightarrow nd$ 
(Ref.\cite{weber89})
and 
$\pi^+ t\rightarrow pd$
(Ref. \cite{sal92})
have been included.  
The data at 450 and 500 MeV were taken from Ref.
\cite{cam87}
using charge independence,
and finally, at 605 MeV, we considered the older data
of Ref.\cite{asl77}.
In terms of integrated cross--section,
that last set of angular data corresponds in Figure 2 
to the datum at $\eta$ = 2.4, which has the largest error bar.
Because of this, the large multiplicative factors
we found at this energy (5.956 for Bonn {\it B}, 5.413 for 
Bonn {\it A}, and 7.982 
for Paris) seem to be attributable
more to normalization problems in the data,
than to the model calculation.
For all the remaining energies, the normalization factors were 
all well around one.

Table 2 compares for the three NN potentials
the integrated cross--section calculated under various conditions.
The first row exhibits the contribution arising
from the sole $\pi$NN vertex (denoted IA, Impulse Approximation).
In the second row (PWa), we show a plane--wave calculation which 
includes the isobar degrees of freedom via the
$\pi$N$\Delta$ vertex.
In the third row, PWb, the number of intermediate NN states
have been increased from 4 up to 18 two--body partial waves,
and from the sole $S$ state 
up to the $D$ states in the $\Delta$N orbital momentum.
The effect in the total cross section is about 3\%;
however, as shown in the next figure,
there is a not large but sizeable change in the angular dependence.
In a forthcoming article we will show that certain spin observables like
$A_{y0}$ are extremely sensitive to the number of waves 
included in the intermediate states.

Finally, in the last two rows (ISIa and ISIb)
the effect of 3--nucleon dynamics in the incoming channel
has been taken into account with the Faddeev formalism. 
Similarly than in the previous two rows,
these differ between each other
for the number of intermediate states.
In both cases, the effect of the three--nucleon dynamics
yields an increase of about 4\% or less.
Around the same value for $\eta$, the $pd\rightarrow \pi^+ t$
cross--section extracted from Ref.
\cite{cam87} gives 35 $\pm$ 3.2 $\mu b$ at $\eta=1.32$,
while data extraction from \cite{weber89} 
gives 32.7 $\pm$ 8.0 at $\eta=1.43$.

Figure 4 shows the differential cross--section calculated
for $\eta=1.36$
in various conditions, in comparison to
the data obtained for $\pi^o$ production 
at 350 MeV assuming charge independence
\cite{cam87}.
The solid line is the solution of the Faddeev equation
for the three--nucleon dynamics 
in the incoming channel.
The corresponding plane--wave results are shown by the dotted line.
The comparison shows that the three--nucleon
dynamics have a relatively small
effect on the differential cross--section.
On a linear scale, the effect is more pronounced at forward 
angles, however, on a log scale it becomes evident
that the overall effect is simply a rescaling of the curve,
without changing the angular dependence.
The dashed line shows the results obtained
with a limited number of channels (a total of 82 instead of 464).
Differences are seen at both forward and backward angles.
It is therefore important
to consider convergence with respect to
the number of three--body states included in the calculation.
Finally, the dashed--dotted line contains the effect
of the sole nucleonic intermediate states.
For reasons of visibility, 
these non--isobaric effects have been multiplied by a factor of 10.
Overall the effect is small, but at backward angles its contribution
is larger than 10\%.
The calculations shown in the figure were performed with
the Paris interaction.
Practically identical angular dependences have been obtained with the 
Bonn {\it B} and Bonn {\it A} potentials.

Finally, Figure 5 exhibits the dependence of the differential cross 
section upon the parameter $\eta$ for forward and backward angles.
Due to the smallness of the three--body effects, 
the calculations have been 
performed in plane--wave approximation.
The theoretical results have been divided by 2
for comparison with the experimental data for $\pi^o$ production
obtained at Saclay \cite{may86}.
The Bonn {\it B} and Paris curves have been normalized
at E = 350 MeV to the data of Ref. 
\cite{cam87}.

\section{Summary and conclusions}

We have analyzed the $pd\rightarrow \pi^+ t$ reaction
around the $\Delta$--resonance with a model calculation
which explicitly includes isobar degrees of freedom and meson--exchange
diagrams. The elementary production/absorption mechanisms were tested
on the simpler $pp\rightarrow\pi^+ d$ reaction. In particular,
the position and width of the isobar resonance were modeled
to reproduce the excitation function of the
$pp\rightarrow\pi^+ d$ process, and off--shell effects
in the baryonic coordinates were taken into account in both $\pi$NN and 
$\pi\Delta$N vertices.
The same absorption mechanisms (without further changes) have been
embedded in a Faddeev--based treatment for the $pd\rightarrow
\pi^+t$ process, where the three--nucleon bound state and the 
three--nucleon continuum dynamics in the initial channel have been 
calculated using the Faddeev--AGS formalism. The computational method is 
similar to the one employed recently for the triton
photodisintegration.
We have checked our results against 
integral and differential cross--section data in the resonance region, 
finding that the unpolarized experimental data
are reproduced reasonably well.

As for the excitation function, we have found that the resonance peak is 
reproduced within the errors without changing the isobar parameters in 
passing from $pp$ pionproduction to the $pd$ one.
We have also found that the magnitude of the curve ({\it i.e.}
the normalization of the cross section) is sensitive
to the nucleon--nucleon interaction used as input for the 3N
bound state wavefunction, with a 25\% difference
between the Bonn potentials and the Paris one.
This sensitivity is larger by a factor of 5 with respect to the 
$pp\rightarrow\pi^+d$ process.The results with the Paris potential
were always smaller than those obtained with the Bonn potentials.

We have then analyzed the angular dependence of the differential 
cross--section in the energy region spread around the resonance peak.
At the peak there are practically no differences in the angular 
distribution with respect to the model selected 
for the nucleon--nucleon interaction. 
In moving away from the peak, the angular distributions begin
to show some interaction dependence: the differences are larger
at higher energies and backward angles. The calculation
with the Paris potential seems slightly in better agreement
with the general trend of the data at backward angles. 
Definitive conclusions, however,
should be drawn only after the $\pi$N interactions in $s$--wave
will be also included.

At the resonance peak we have singled out the role
of non isobaric $\pi$N interaction in $p$--wave
(the IA term), analyzed the convergence of the calculation
with respect to the number of intermediate three--baryon 
partial waves included, and considered the effect of the three--body 
dynamics in the nucleon--deuteron channel.
It turns out that the size of these effects are all comparable.
Indeed, the contribution of the IA term 
with respect to the total cross section
ranges from 2 to 4 \% depending on 
the model interaction which has been used.
Differences of the same size are found when the calculation
performed with a strictly necessary set of partial waves
is compared with converged results. Finally,
by means of the Faddeev--AGS formalism,
it was possible to ascribe a 4 \% effect 
to the contribution due to the three--nucleon dynamics
in the nucleon--deuteron channel.
It is obvious that this is equivalent to say
that for the inverse pionabsorption reaction,
final state interactions contribute with the
same 4\% amount.

In conclusion, we have here confirmed the expectations, 
previously formulated in Refs. \cite{casveca,catcan},
that a careful embedding of the basic pion absorption/production
matrix elements in a Faddeev--based treatment of the few--nucleon 
dynamics is a very important tool for understanding the
hadronic processes in nuclei at intermediate energies.
Once this point has been settled, it is possible to move
further on with the same approach and tackle other, more refined 
experimental data underlining the pion few--nucleon 
systems.

Such aspects are, {\it e.g.}, the pionic absorption
on the diproton \cite{diproton},
the wealth of experimental data involving polarization phenomena
\cite{falk,cam87,lol82,pol-phen}, the pion--induced reaction at energies
around the pionic threshold\cite{thre-en}, 
and, finally, the meson--absorption
coincidence experiments ($\pi^+$,pp) at non conjugated angles 
with the connected puzzle of the ``genuine" three--body mechanisms
\cite{3B-reviews}.
However, before that all these aspects can be theoretically
disentangled,
an important improvement is needed in our treatment
(as well as in any other approach).
This is connected with the role played by the pion--nucleon 
$s$--wave interaction (in both its isoscalar and isovector
components) in multiple rescattering processes. Stated in other
words, it is the role played by pion--nucleus final 
state interaction (or initial, depending on 
the selected direction in time). This aspect
is still missing in the present treatment
of the d(p,$\pi^+$)t process, but a flavour of its importance
can be immediately percieved in the simpler
case of the p(p,$\pi^+$)d reaction,
by glancing at the
dashed and dotted--dashed curves passing through the data
in Figure 1.

\begin{acknowledgments}

We thank G. Cattapan, P.J. Dortmans, G. Pisent, and J.P. Svenne
for scientific discussions and interest at an early stage of this
research project.
L.C. wishes to acknowledge helpful discussions and correspondence with 
K. Amos, Q. Ingram, B. Mayer, J. K\"ohler, P. Weber,  A. Lehmann,  W. Falk, 
and D. Hutcheon.
W. S. wishes to thank the warm hospitality of the University of Padova
during several visits.
The work of W. S. was supported by the INFN and the
Deutsche Forschungsgemeinschaft under Grant No. Sa 327/23-1.
\end{acknowledgments}

\vfill
\eject

\setcounter{figure}{0}
\begin{figure}
FIG. 1. Excitation function for $\pi^+$ $d$ production
(in microbarn) from $p p$ collisions. The parameter $\eta$ 
corresponds to the pion momentum (c.m.) divided by the pion mass.
The full line is the result obtained with Bonn {\it B} potential
and with $p$--wave pion--nucleon interactions which includes
both isobar and non--isobar degrees of freedom. 
Practically indistinguishable from the full line,
there is a dotted line which correspond to a similar calculation
but with Paris potential.
The other two curves (dashed, and dashed--dotted, respectively)
correspond to the inclusion of pion--nucleon
initial state interaction in $s$--wave and differ between each other
for the NN potential being used. The Bonn {\it B} in the dashed case, Paris 
in the dotted--dashed case. 
\label{figuno}
\end{figure}

\begin{figure}
FIG. 2. Excitation function of pion production  
via proton--deuterium collisions (in microbarn). The 
calculation includes also $\Delta$ excitation 
and has been performed with the same parameters
as in Figure 1.
\label{figdue}
\end{figure}

\begin{figure}
FIG. 3. Differental cross--section 
for $\pi^+$ production in proton--deuterium collision.
\label{figtre}
\end{figure}

\begin{figure}
FIG. 4. $\pi^+$--production differential cross--section
at $\eta=1.36$. The full line shows
the results with inclusion of nucleon--nucleon
initial--state interactions.
The dotted line represents the plane--wave calculation.
The dashed line shows the results with a
limited number of channels. The dashed--dotted line
shows the contribution of the sole non--isobar degrees
of freedom, magnified by a factor 10. 
All calculations were performed with the Paris 
potential.
\label{figquattro}
\end{figure}

\begin{figure}
FIG. 5. Unpolarized differential cross--section for $\pi^o$ 
production in proton--deuterium collisions. Circles and triangles
represent measurments taken at Saturne \cite{may86} 
at forward and backward (c.m.)
angles, respectively. 
The solid and broken curves are forward--angle 
calculations with Bonn {\it B} and Paris potentials, respectively,
the dashed and dotted-dashed curves are the corresponding
backward--angle results.
\label{figcinque}
\end{figure}

\newpage

\begin{table}[ht]
\begin{center}
\begin{tabular}{|c|c|c|c|c|c|c|c|}
  $J^P= {1 \over 2}^-$ &$J^P= {1 \over 2}^+$ &$J^P= {3 \over 2}^-$ &
  $J^P= {3 \over 2}^+$ &$J^P= {5 \over 2}^-$ &$J^P= {5 \over 2}^+$ &
  $J^P= {7 \over 2}^-$ &$J^P= {7 \over 2}^+$\\
\hline
$^1S_0P_{1\over 2}$&$^1S_0S_{1\over 2}$&$^3S_1P_{1\over 2}$&$^3S_1S_{1\over 2}
$&$^1D_2P_{1\over 2}$&$^1D_2S_{1\over 2}$&$^1D_2P_{3\over 2}$
&$^1D_2D_{3\over 2}$\\
$^3S_1P_{1\over 2}$&$^3S_1S_{1\over 2}$&$^3D_1P_{1\over 2}$&$^3D_1S_{1\over 2}
$&$^3S_1P_{3\over 2}$&$^3S_1D_{3\over 2}$&$^3S_1F_{5\over 2}$
&$^3S_1D_{5\over 2}$\\
$^3D_1P_{1\over 2}$&$^3D_1S_{1\over 2}$&$^1D_2P_{1\over 2}$&$^1D_2S_{1\over 2}
$&$^3D_1P_{3\over 2}$&$^3D_1D_{3\over 2}$&$^3D_1F_{5\over 2}$
&$^3D_1D_{5\over 2}$\\
$^3S_1P_{3\over 2}$&$^3S_1D_{3\over 2}$&$^1S_0P_{3\over 2}$&$^1S_0D_{3\over 2}
$&$^1D_2F_{3\over 2}$&$^1D_2D_{3\over 2}$&$^1D_2F_{5\over 2}$
&$^1D_2D_{5\over 2}$\\
$^3D_1P_{3\over 2}$&$^3D_1D_{3\over 2}$&$^3S_1P_{3\over 2}$&$^3S_1D_{3\over 2}
$&$^1S_0F_{5\over 2}$&$^1S_0D_{5\over 2}$&$^1S_0F_{7\over 2}$
&$^1S_0G_{7\over 2}$\\
$^1D_2P_{3\over 2}$&$^1D_2D_{3\over 2}$&$^3D_1P_{3\over 2}$&$^3D_1D_{3\over 2}
$&$^3S_1F_{5\over 2}$&$^3S_1D_{5\over 2}$&$^3S_1F_{7\over 2}$
&$^3S_1G_{7\over 2}$\\
$^1D_2F_{5\over 2}$&$^1D_2D_{5\over 2}$&$^1D_2P_{3\over 2}$&$^1D_2D_{3\over 2}
$&$^3D_1F_{5\over 2}$&$^3D_1D_{5\over 2}$&$^3D_1F_{7\over 2}$
&$^3D_1G_{7\over 2}$\\
&&$^3S_1F_{5\over 2}$&$^3S_1D_{5\over 2}
$&$^1D_2F_{5\over 2}$&$^1D_2D_{5\over 2}$&$^1D_2F_{7\over 2}$
&$^1D_2G_{7\over 2}$\\
&&$^3D_1F_{5\over 2}$&$^3D_1D_{5\over 2}
$&$^3S_1F_{7\over 2}$&$^3S_1G_{7\over 2}$&$^3S_1H_{9\over 2}$
&$^3S_1G_{9\over 2}$\\
&&$^1D_2F_{5\over 2}$&$^1D_2D_{5\over 2}
$&$^3D_1F_{7\over 2}$&$^3D_1G_{7\over 2}$&$^3D_1H_{9\over 2}$
&$^3D_1G_{9\over 2}$\\
&&$^1D_2F_{7\over 2}$&$^1D_2G_{7\over 2}$
&$^1D_2F_{7\over 2}$&$^1D_2G_{7\over 2}$&$^1D_2H_{9\over 2}$
&$^1D_2G_{9\over 2}$\\
&&&&$^1D_2G_{9\over 2}$&$^1D_2H_{9\over 2}$&&
\end{tabular}
\end{center}
\caption{Example of three--nucleon partial waves 
included in our calculation. The notation is 
${^{2 s+1}{ l}_j}\ { \lambda}_I$,
where $s$, $l$, are $j$ are spin, orbital, and
total momentum of the pair, while
$\lambda$ and $I$ are orbital and total momentum for the
spectator nucleon.
This set of 82 states corresponds to the inclusion of 4
two--nucleon states.  In our actual calculation
464 three--body partial waves have been considered,
corresponding to the inclusion of 18 two--nucleon states.}
\end{table}

\vspace{2cm}

\begin{table}[ht]
\begin{center}
\begin{tabular}{cccc} 
$\ $ 
& Bonn {\it A} 
& Bonn {\it B} 
& Paris \\
\hline
IA 
& 0.58  
& 0.72  
& 0.93 \\
PWa 
& 30.1  
& 28.0  
& 21.7 \\
PWb 
& 30.7  
& 28.7  
& 22.5 \\
ISIa  
& 31.3  
& 29.0  
& 22.5 \\
ISIb  
& 32.0  
& 29.9  
& 23.4 \\
\end{tabular}
\end{center}
\caption{Calculated production cross sections
(in microbarns) at $\eta$ = 1.36 for various NN potentials. 
The Impulse Approximation (first row)
represents a bare plane--wave calculation 
without $\Delta$ rescatterings.
The second and third rows include also
the $\Delta$ degrees of freedom but use
a plane--wave approximation in the incoming channel.
They differ between each other for the number of partial
waves included in the intermediate states.
In case $a$, 4 two--nucleon states have been included
and only $S$--wave orbital momentum 
for the intermediate $\Delta$--nucleon state has been considered.
In case $b$, 18 two--nucleon states 
and intermediate $\Delta$--nucleon states in $S$, $P$, and $D$
waves have been taken into account.
In the last two rows, where the same number of partial waves
have been included as in the previous two, the $p$--$d$ 
interactions in the initial state have been included via a 
Faddeev calculation.}
\end{table}

\begin{table}[h]
\begin{tabular}{lcccc}
 partial-wave       & PEST & BAEST & BBEST \\
\hline                                    
 $^1S_0$            &   5  &    5   &   5  \\
 $^3S_1-\,^3\!D_1$  &   6  &    6   &   6  \\
 $^1D_2$            &   5  &    4   &   4  \\
 $^3D_2$            &   5  &    4   &   4  \\
 $^1P_1$            &   5  &    4   &   4  \\
 $^3P_1$            &   5  &    4   &   4  \\
 $^3P_0$            &   5  &    4   &   4  \\
 $^3P_2-\,^3\!F_2$  &   7  &    5   &   5  \\
\end{tabular}
\caption{Ranks of the two-body partial-waves of the Paris, Bonn {\it A},
and Bonn {\it B} potentials in the EST-representation.
\label{ranktab}}
\end{table}

\begin{table}[h]
\begin{center}
\begin{tabular}{ccc} 
 BAEST & BBEST  & PEST \\
\hline
 -8.284 & -8.088 & -7.3688 
\end{tabular}
\caption{Calculated binding energies for the triton.
The total angular momentum of the two--body subsystem
was restricted to $j\leq 2$.
This led to 18 three--body channels in the channel--spin 
coupling.\label{energytab}}
\end{center}
\end{table}

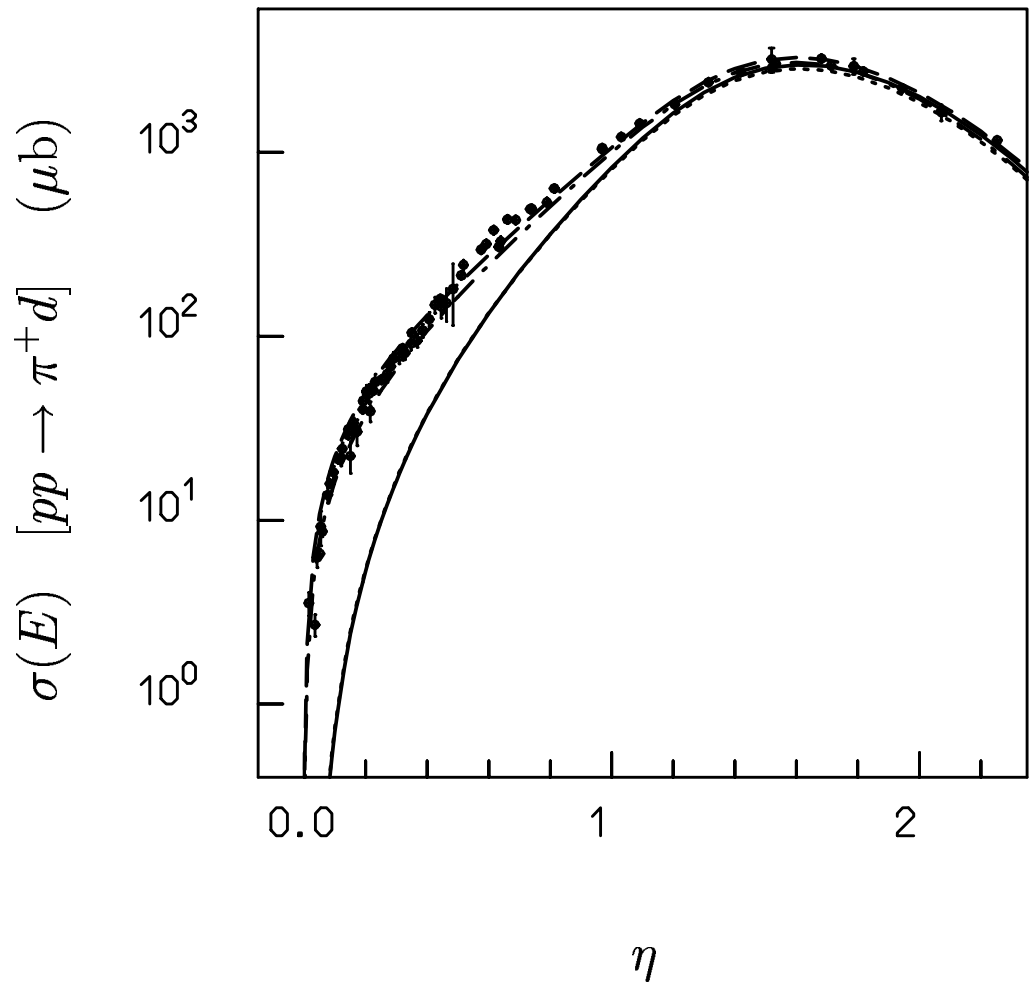
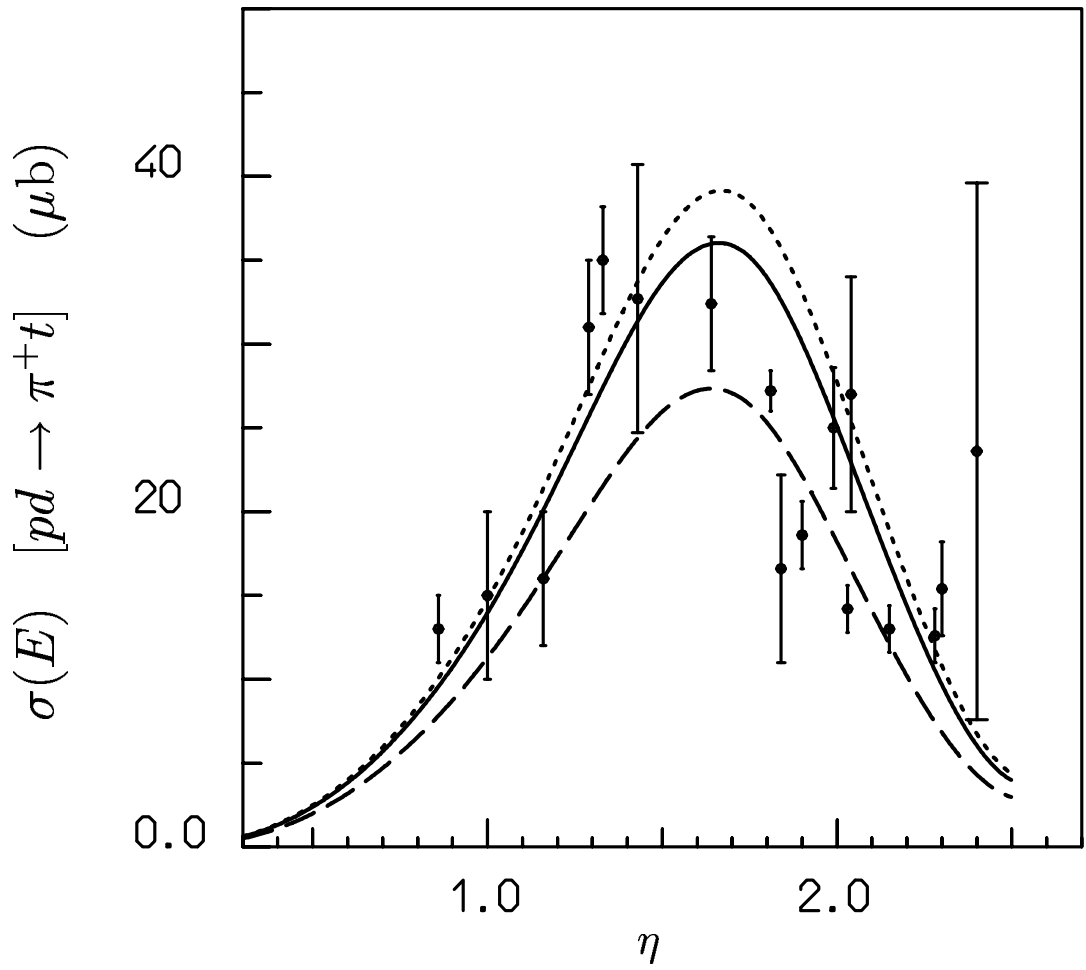
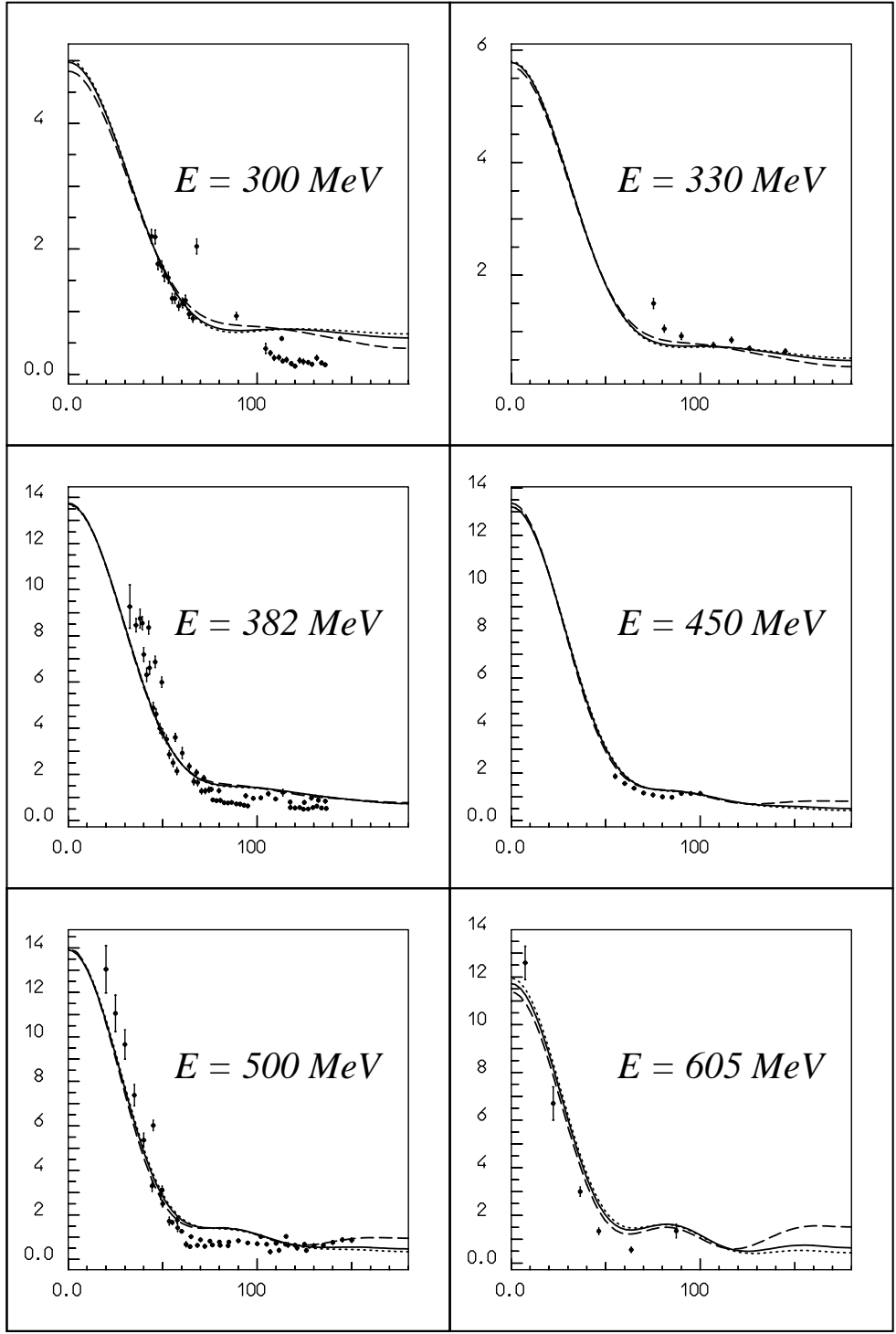
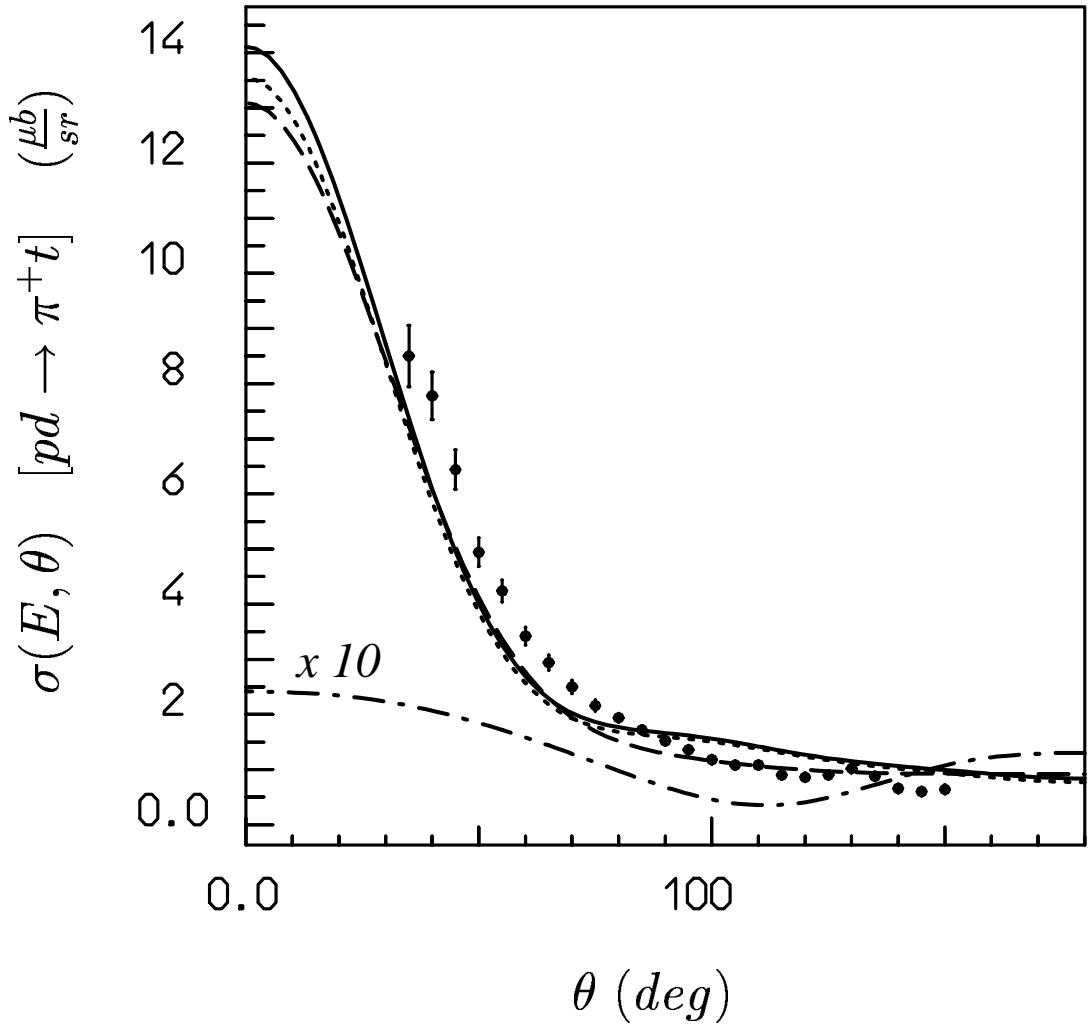
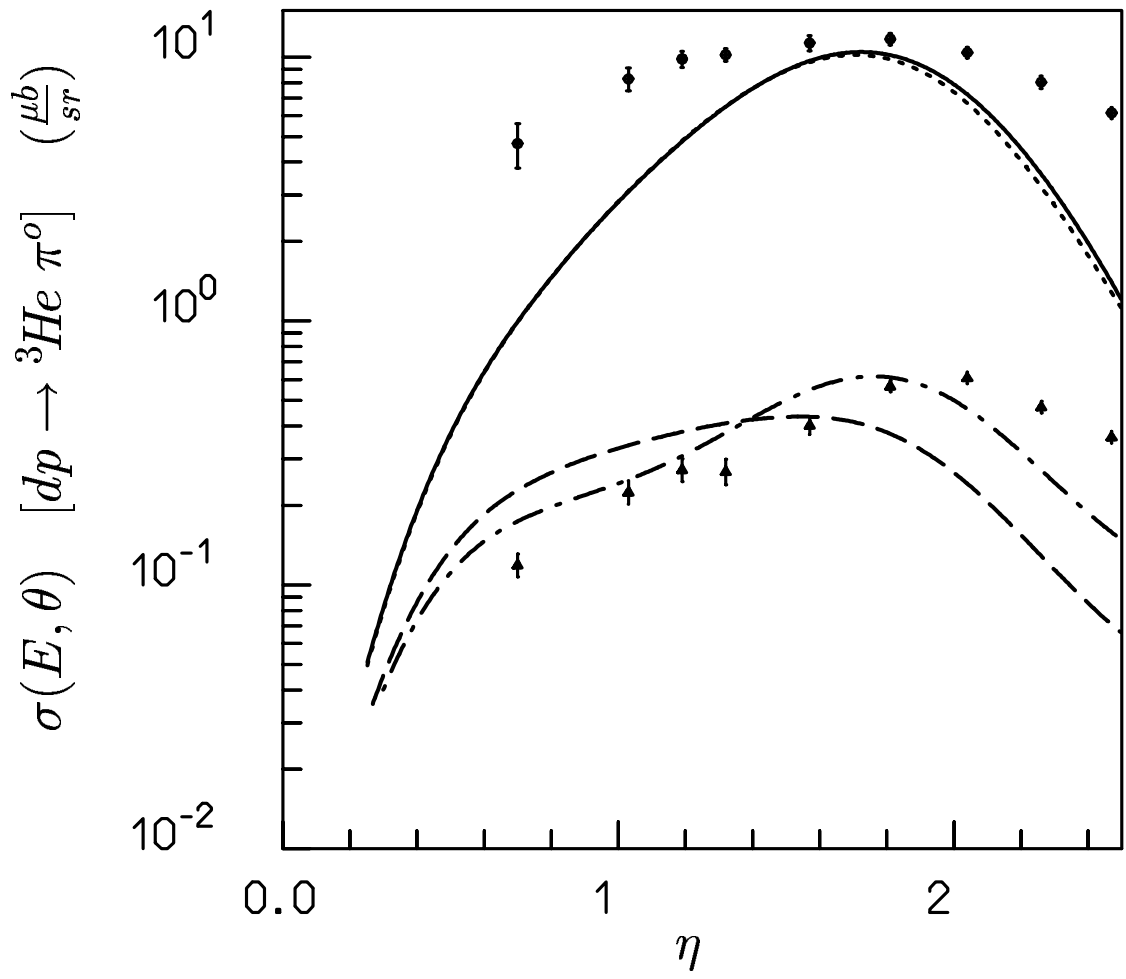
\end{document}